\documentclass[twoside,leqno,twocolumn,hidelinks]{article}

\usepackage[letterpaper]{geometry}

\usepackage{hyperref}
\usepackage{setspace}

\usepackage{amsmath}
\usepackage{amsfonts}
\usepackage{amssymb}
\usepackage{balance}

\usepackage{cleveref}

\usepackage[ruled,vlined,algo2e]{algorithm2e}

\DontPrintSemicolon
\SetKwComment{tcp}{$\triangleright$ }{}
\SetVlineSkip{0cm}
\SetKwInOut{Param}{Param}
\SetKwInOut{Input}{Input}

\usepackage{graphicx}
\usepackage{wrapfig}
\usepackage{subfigure}
\usepackage{colortbl}
\usepackage{multicol}
\usepackage{multirow}

\renewcommand{\mid}{:}

\usepackage{xargs}
\usepackage[pdftex,dvipsnames]{xcolor}  
\usepackage[colorinlistoftodos,prependcaption,textsize=scriptsize]{todonotes}

\newcommandx{\uvc}[2][1=]{\todo[color=magenta!50,#1]{\sf \textbf{\"Umit:} #2}\xspace}

\definecolor{darkgreen}{rgb}{0,0.4,0}

\DeclareMathOperator*{\argmin}{argmin}

\newcommandx{\mLI}{\ensuremath{\text {\sc mLI}}\xspace}
\newcommandx{\mNC}{\ensuremath{\text {\sc mNC}}\xspace}

\newcommand\supersmallfont{\fontsize{7}{8}\selectfont}

\usepackage{ltexpprt}

\begin{document}

\title{Heuristics for Symmetric Rectilinear Matrix Partitioning}

\author{
  Abdurrahman Ya\c{s}ar
  \thanks{School of Computational Science and Engineering, Georgia Institute
  of Technology, Atlanta, GA 30332
  (\href{mailto:ayasar@gatech.edu}{ayasar@gatech.edu},
   \href{mailto:umit@gatech.edu}{umit@gatech.edu})}
\and
  \"{U}mit V. \c{C}ataly\"{u}rek\footnotemark[1]
}

\date{}

\maketitle

\fancyfoot[R]{\scriptsize{Copyright \textcopyright\ 20XX by SIAM\\
Unauthorized reproduction of this article is prohibited}}

\begin{abstract} \small\baselineskip=9pt
Partitioning sparse matrices and graphs is a common and important problem
that arises in
many scientific and graph analytics applications. In this work, we are
concerned with a spatial partitioning called {\em rectilinear partitioning}
(also known as generalized block
distribution) of sparse matrices, which is necessary for {\em tiled} (or {\em
blocked}) execution of sparse matrix and graph analytics kernels. More
specifically, in this work, we address the problem of {\em symmetric} rectilinear
partitioning of square matrices. By symmetric, we mean having the same
partition on rows and columns of the matrix, yielding a special tiling where
the diagonal tiles (blocks) will be squares. We propose five heuristics to
solve two different variants of this problem, and present a thorough
experimental evaluation showing the effectiveness of the proposed
algorithms.
\end{abstract}

\section{Introduction}
\label{sec:intro}

After advances in the social networks and the rise of interactions on the web,
we are witnessing an enormous growth in the volume of generated data. A large
portion of this data remains sparse and irregular. Graphs and sparse matrices
are used to store and analyze an important portion of this data. However,
analyzing data stored in that kind of irregular data structures is becoming
more and more challenging, especially for traditional architectures due to the
growing size of these irregular problems.
The sheer size of the problems necessitates parallel execution. There
have been many studies developing parallel graph and sparse matrix algorithms for
shared and distributed memory systems as well as GPUs and hybrid systems.
Effective data and computation partitioning is the
first step to propose efficient portable (parallel) algorithms~\cite{Gill18-VLDB}.

Two-dimensional matrix partitioning is a hard problem and has been used in
dense linear algebra~\cite{Oleary85-ACMC} for a long time.
Checkerboard partitioning, where the partitioned matrix maps naturally onto a
2D mesh of processors, is widely used in earlier two-dimensional matrix
partitioning~\cite{Hendrickson950IJHSC,Lewis94-HPCC}. Checkerboard partitioning
is highly useful to limit the total
number of messages on distributed settings. However, these works are suited to
dense or well structured sparse matrices.

In the context of this paper, we focus on spatial, two-dimensional
checkerboard-like partitioning problem that we call {\em Symmetric
Rectilinear Partitioning}. Here, we assume that given matrix is square and
we would like to partition that matrix into $p \times p$ tiles such that
by definition diagonal blocks will be squares. This type of partitioning is
very convenient if one wants to gather information along the rows/columns
and distribute along columns/rows. Also, in the context of graphs, diagonal
tiles can be visualized as sub-graphs and any other tile represents the edges
between two sub-graphs. This type of partitioning becomes highly
useful to reason about graph algorithms.
For example, in a concurrent work, we have leveraged the
symmetric rectilinear partitioning for developing a block-based triangle counting
algorithm~\cite{Yasar19-HPEC} that reduces data movement during, both sequential and parallel,
execution and also naturally suitable for heterogeneous architectures.

In this work, we define two variants of the symmetric rectilinear partitioning
problem and we propose refinement based and probe-based partitioning heuristics
to solve these problems. Refinement based
heuristics~\cite{Manne96-IWAPC, Nicol94-JPDC} apply a dimension reduction
technique to map the two-dimensional problem into one-dimension and compute partition
vector on one-dimensional data by running an optimal partitioning
algorithm~\cite{Pinar04-JPDC}. Probe-based algorithms compute the partitioning
vector by seeking for the best cut for each point. The novelty of the proposed
approaches is to use the natural order of the matrix instead of running expensive
hypergraph models or graph partitioning algorithms to order vertices. We combine
lightweight spatial partitioning techniques with simple heuristics.

Contributions of this work are as follows:

\begin{itemize}
\item We propose heuristics for symmetric rectilinear partitioning
problem that does not require row orderings.

\item We evaluate the effect of the simple, degree-based and
RCM based vertex orderings on the tile distributions.

\item We experimentally evaluate the performances of proposed
algorithms wrt. state-of-the-art algorithms in different settings.
\end{itemize}

Our experimental results show that our proposed algorithms are
very effective in finding symmetric rectilinear partitions. In all instances,
our algorithms produce similar or better load-imbalanced solutions than
Nicol's~\cite{Nicol94-JPDC} rectilinear partitioning algorithm, which has more
freedom in choosing row and column partitions.

\section{Problem Definition}
\label{sec:pdef}

\begin{figure}[ht]
  \centering
  \subfigure[Regular: $C_c=\{0,5,8,10\}$, $C_r=\{0,4,8,10\}$]{\includegraphics[width=.48\linewidth]{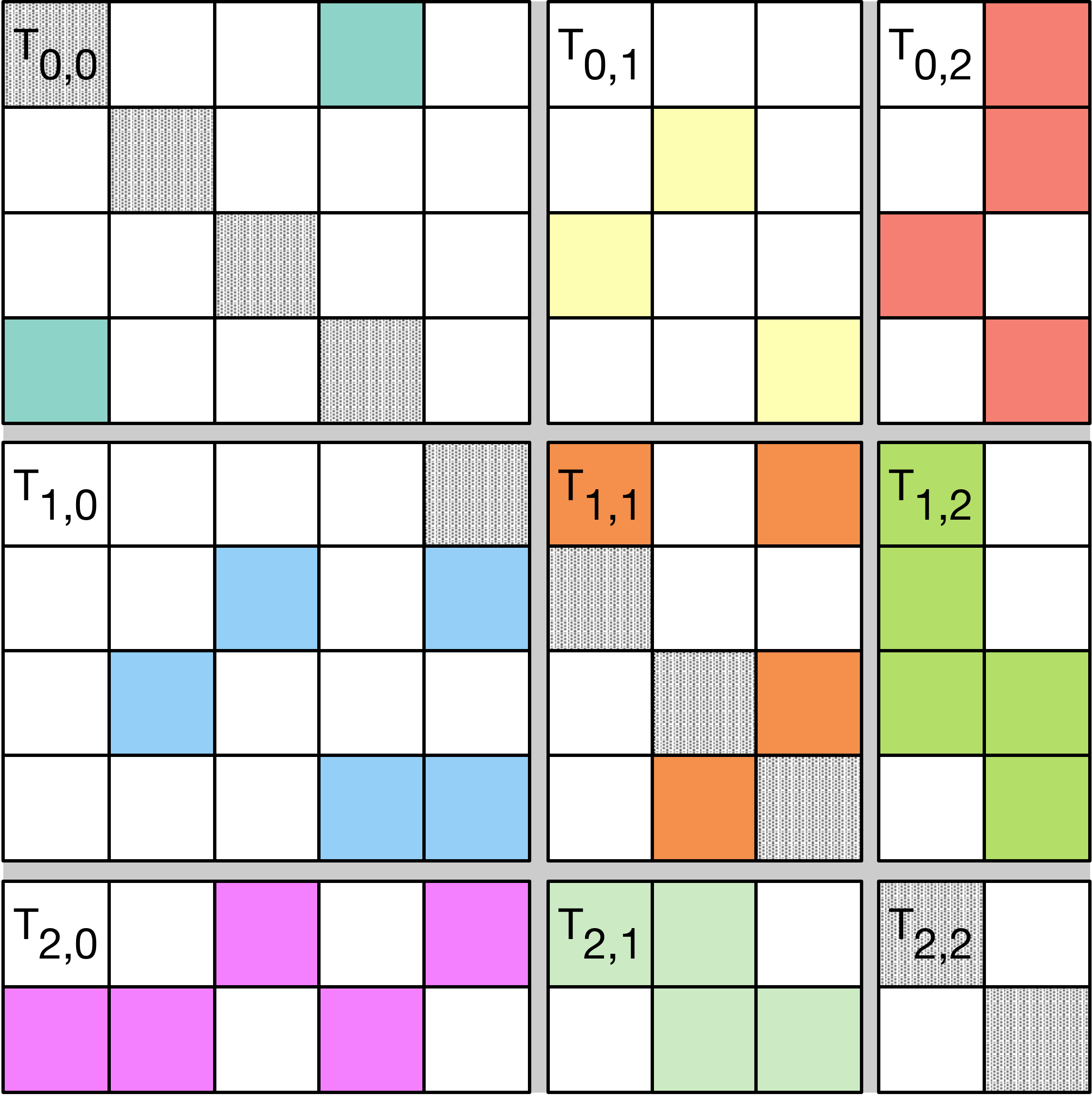}
  \label{fig:ex-nic}}
  \subfigure[Symmetric: $C_c=C_r =\{0,5,8,10\}$]{\includegraphics[width=.48\linewidth]{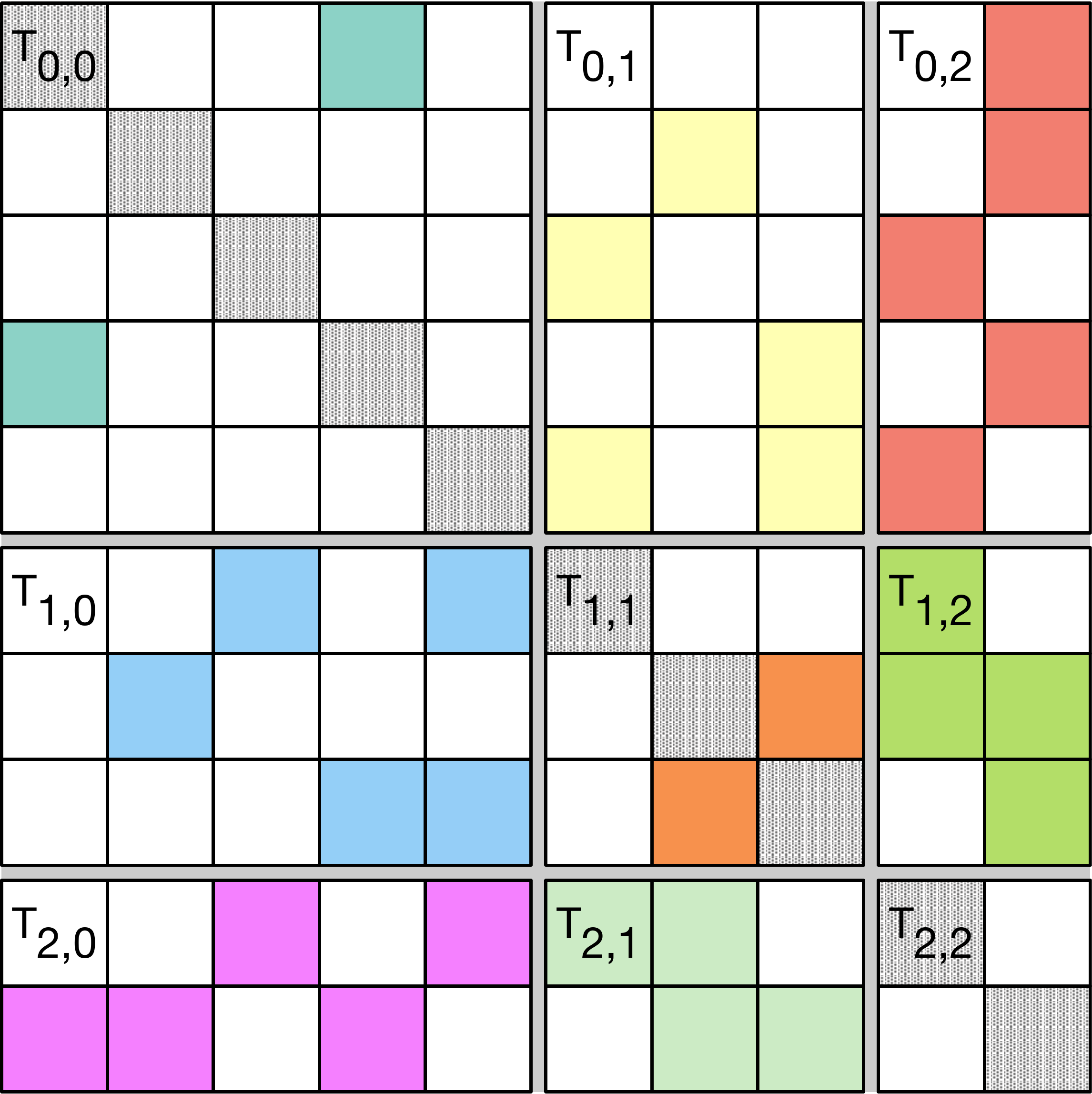}
  \label{fig:ex-ptc}}
  \caption{$3 \times 3$ Rectilinear Partitioning}
  \label{fig:exsr}
\end{figure}

In this paper, we are concerned with partitioning sparse matrices.
In particular, we are interested in partitioning adjacency matrix
representation of graphs. A directed graph $G=(V, E)$, consists of a
set of vertices $V$ and a set of edges $E$. A directed edge $e$ is
referred as $e = (u, v) \in E$, where $u, v \in V$, and $u$ is called
the source of the edge and $v$ is called the target. The neighbor list
of a vertex $u \in V$ is defined as $N[u] = \{v\in V\mid(u, v)\in E\}$.
We will use $n$ and $m$ for number of vertices and edges, respectively,
i.e., $n=|V|$ and $m=|E|$. Let $A_G$ be the adjacency matrix representation
of the graph $G$, where all edges are represented with nonzeros, and the
rest of the entries will be zero. That is, $A_G$ is an $n\times n$ matrix,
where $\forall (u, v) \in E$, $A_G[u,v] =1$, and everything else will be 0.
Without loss of generality, we will assume source vertices are represented
as rows, and target vertices represented as columns. In other words,
elements of $N[u]$ will correspond to column indices of nonzero elements
in row $u$. We will also simply refer to matrix $A_G$ as $A$ when $G$ is
clear in the context. \Cref{tab:notation} lists the notations used in
this paper.

\begin{table}[tbp]
\begin{small}
\begin{tabular}{r  l}
\textbf{Symbol} & \textbf{Description}                \\
\hline
$G=(V, E)$      & A directed graph $G$ with vertex \\
                & \ and edge sets, $V$ and $E$, respectively \\
$n=|V|$         & number of vertices \\
$m=|E|$         & number of edges \\
$A_G$           & $n\times n$ adjacency matrix of $G$ \\
$N[u]$          & Neighbor list of vertex $u$\\
$C$             & Partition vector; $C=\{c_0, \dots, c_p\}$ \\
$C_c$, $C_r$    & Column and row partition vectors\\
$T_{i,j}$       & Tile $i,j$ \\
$\lambda(A, C_c, C_r)$    & Load imbalance for partition vectors\\
$\lambda(A, C_c, C_r, k)$ & Load imbalance among $T_{i,j}$ st. $i,j \leq k$\\
\hline
\end{tabular}
\end{small}
\caption{Notations used in this paper.}
\label{tab:notation}
\end{table}

Given adjacency matrix of $G$, $A_G \in \mathbb{B}^{n \times n}$ and an
integer $p$, $1 \leq p \leq n$. Let $C$ be a partition vector
that consists of sequence of $p+1$ integers such that
$0=c_0<c_1<\dots<c_p=n$. Then $C$ defines a
partition of $[0, n]$ into $p$ intervals
$[c_i, c_{i+1}-1]$ for $0 \leq i \leq p-1$.

\begin{Definition}{Rectilinear Partitioning.}
Given $A$, and two integers, $p$ and $q$, a rectilinear partitioning
consists of a partition of $[0, n]$ into $p$ intervals
($C_c$, for columns) and into $q$ intervals ($C_r$, for rows) such that $A$ is partitioned into
non-overlapping $p \times q$ contiguous tiles.
\end{Definition}

In rectilinear partitioning, a column partition vector, $C_c$, and a row partition vector, $C_r$,
together generate $p \times q$ tiles.
For $i \in [0,p]$ and $j\in[0,q]$, we denote $(i,j)$-th tile by $T_{i,j}$ and
$|T_{i,j}|$ denotes the number of nonzeros in $T_{i,j}$.
Given partition vectors, quality of partitioning can be defined using
load imbalance, $\lambda$, among the tiles, which is computed as
$$\lambda(A, C_c, C_r) = \frac{L_{max}}{L_{avg}} - 1$$
\noindent where
$$L_{max} = \max_{0\leq i,j \leq n}\{ |T_{i,j}| \}$$ and
$$L_{avg}=\frac{\sum_{0 \leq i,j \leq n}|T_{i,j}|}{p \times q} =
\frac{2\times m}{p \times q}.$$
A solution which is perfectly balanced achieves a load imbalance,
$\lambda$, of $0$. \Cref{fig:ex-nic} presents a toy example for
rectilinear partitioning where $C_c=\{0,5,8,10\}$, $C_r=\{0,4,8,10\}$
and $\lambda(A, C_c, C_r) = \frac{5}{3.6}-1 = 0.39$.

\begin{Definition}{Symmetric Rectilinear Partitioning.}
Given $A$ and $p$, symmetric rectilinear partitioning
can be defined as partitioning $[0, n]$ into $p$ intervals such that
$A$ partitions into $p \times p$ non-overlapping contiguous tiles where
diagonal tiles are squares.
\end{Definition}

In symmetric rectilinear partitioning, same partition vector, $C_c=C_r$,
is used for row and column partitioning.
\Cref{fig:ex-ptc} presents a toy example for the symmetric rectilinear
partitioning  where $C_c= C_r = \{0,5,8,10\}$
and $\lambda(A, C_c, C_r) = \frac{5}{3.6}-1 = 0.39$

In the context of this work, we consider two symmetric rectilinear
partitioning problems. The first problem ({\sc minLoadImbal}, or \mLI in short),
consists in finding the optimal partition vector, $C$, that minimizes the load
imbalance, for a given the matrix $A$ and an integer $p$:

\begin{align*}
\mLI( A, p ) &= \argmin_{C} \lambda(A, C, C)
\end{align*}

The second problem ({\sc minNumCuts}, or \mNC in short), is the dual
of the first problem. For a given a matrix, $A$, and an integer, $Z$,
this problem consists of finding the minimum number of intervals, $p$,
that will partition the matrix $A$, where the number of nonzeros in all
tiles is bounded by $Z$.

\begin{align*}
\mNC( A, Z ) &= \argmin_{p\in [1, n]} \lambda(A, \mLI( A, p ), \mLI( A, p )) \\
            & \leq \frac{Z\times p^2}{ m}
\end{align*}

\section{Related Work}
\label{sec:related}

Two-dimensional matrix distributions have been widely used in dense
linear algebra. Most of these distributions are
cartesian~\cite{Hendrickson950IJHSC}; where the same partitioning
vector is used for row and column partition. For sparse and irregular
problems finding a good vector that can be aligned with both dimensions
is even harder. Therefore, many non-cartesian two-dimensional matrix
partitioning methods have been
proposed~\cite{Manne96-IWAPC, Nicol94-JPDC, Saule12-JPDC-spart} for
sparse and irregular problems.

One way to overcome the hardness of proposing one partition vector for rows and
columns is to propose a partition vector for rows and columns.
This problem is named as rectilinear partitioning~\cite{Nicol94-JPDC}
(or generalized block distribution~\cite{Manne96-IWAPC}).
Independently, Nicol~\cite{Nicol94-JPDC}
and Manne and S{\o}revik~\cite{Manne96-IWAPC} proposed an algorithm to solve this
problem that is based on iteratively improving a given solution by alternating
between row and column partitioning.
These algorithms
transform two dimensional (2D) rectilinear partitioning problem into
one-dimensional (1D) partitioning problem using a heuristic
and iteratively improves the solution.
The one-dimensional partitioning problem is built by
setting the load of an interval of the problem as the maximum of the load of
the interval inside each stripe of the fixed dimension. This refinement
technique is presented in Algorithm~\ref{alg:rowparti}.
Here, {\sc optimal1DPartition}($P$) is a function that returns the optimal 1D partition
(which is also known as {\em chains-on-chains} partitioning)~\cite{Pinar04-JPDC}.
Hence, Algorithm~\ref{alg:rowparti} returns the optimal 1D row partition for
the given column partition $C_c$.

\begin{algorithm2e}[t]
\begin{small}
  \tcp*[l]{Array to store max of interval sums for each vertex}
  $P[i] = 0$, for $ 0 \leq i \leq n+1$ \;

  \tcp*[l]{for each row}
  \For{$i=0$ to $n-1$}{
    \tcp*[l]{Array to store interval sums}
    $nnz[k] = 0$, for $0 \leq k \leq p$ \;
    \For{{\bf each} $j$, where $A_G[i,j]=1$}{
      $k \leftarrow 0$ \tcp*[l]{Interval index}
      \While{$j \geq C[k+1]$}{
        $k \leftarrow k+1$ \tcp*[l]{Find the interval}
      }
      $nnz[k] \leftarrow nnz[k]+1$ \;
    }
    $P[i+1] \leftarrow \max_k\{nnz[k]\}$\;
  }

  \tcp*[l]{Compute prefix sum}
  \For{$i=1$ to $n$}{
    $P[i] \leftarrow P[i]+P[i-1]$ \;
  }

  \tcp*[l]{Return the output of 1D partitioning}
  \Return \sc{optimal1DPartition($P$)}
  \caption{{\sc refinement($A_G, C_c, p$)}}
  \label{alg:rowparti}
\end{small}
\end{algorithm2e}

Computing the optimal solution was shown to be NP-hard by Grigni and
Manne~\cite{Grigni96-IWPAISP}. In fact, their proof shows that the
problem is NP-hard to approximate within any factor less than $2$. Khanna et
al.~\cite{Khanna97-ICALP} have shown the problem to be constant-factor
approximate.

Rectilinear partitioning may still cause high load-imbalance due to
generalization. Jagged partitions (or Semi Generalized Block
Distribution~\cite{Grigni96-IWPAISP}) tries to overcome this problem by
distinguishing between the main dimension and the auxiliary dimension.
The main dimension is split into $p$ intervals and each of these intervals
partition into
$q$ rectangles in the auxiliary dimension.
Each rectangle of the solution must have its main dimension matching one of
these intervals. The auxiliary dimension of each rectangle is arbitrary.
Saule et al.~\cite{Saule12-JPDC-spart} present multiple variants and
generalization of jagged partitioning.

\section{Symmetric Rectilinear Partitioning}
\label{sec:algs}

We propose five different algorithms for two variants of the symmetric rectilinear
partitioning problem. These algorithms can be classified as refinement
based and probe-based. In this section, we explain how these algorithms
are designed.

\subsection{Heuristics for the {\sc minLoadImbal} problem.}

We propose three algorithms for \mLI problem. Two of these algorithms,
{\em Pick best direction (first) (PBD)} and
{\em Pick best (in each) iteration (PBI)},
leverages previously defined refinement technique (see \Cref{sec:related})
into the symmetric rectilinear partitioning problem.
Note that these two algorithms have no convergence guarantee.
The third algorithm, {\em Probe target cut (PTC)},
implements another heuristic and probes the minimum load imbalance
by moving in partition point in the diagonal of the matrix.

\subsubsection{Pick best direction (first) (PBD)}
\label{subsec:pbd}

algorithm first applies row-based and column-based refinement independently
to simply find the optimal 1D row and column partitions and chooses the one
that gives the best load imbalance. Then, iteratively applies the refinement
algorithm only in this direction until it reaches the iteration limit ($\tau$)
or partition vector does not change significantly (computed using 2-norm).
This procedure is presented in Algorithm~\ref{alg:pbd}.

The primary advantage of this algorithm is its simplicity. This algorithm can be
easily parallelizable like~\cite{Manne96-IWAPC,Nicol94-JPDC}.
However, choosing a direction at the beginning may cause missing information that
can be gathered from the other direction and the solution may converge to a
local optimum quickly.

\begin{algorithm2e}[ht]
\begin{small}
  \tcp*[l]{Current ($C$) and previous ($C'$) partition vectors}
  $C[0] = 0$; $C[j] = n$, for $ 1 \leq j \leq p+1$ \;
  $C'[j] = n$, for $ 0 \leq j \leq p+1$ \;

  \tcp*[l]{Apply 1D partitioning refinement}
  $C_r \leftarrow$ {\sc refinement}($A, C, p$) \tcp*[r]{Row based}
  $C_c \leftarrow$ {\sc refinement}($A^T, C, p$) \tcp*[r]{Column based}

  \tcp*[l]{Aligning same partition vector for rows and columns}
  $L_{r} \leftarrow \lambda(A, C_r, C_r )$ \tcp*[r]{Row based imbalance}
  $L_{c} \leftarrow \lambda(A, C_c, C_c )$ \tcp*[r]{Column based imbalance}
  \eIf{ $L_{r}<L_{c}$ }{
      $C \leftarrow C_r$ \;
    }{
      $C \leftarrow C_c$ \;
    }

  $i \leftarrow 0$ \;
  \While{ $i<\tau$ {\bf and} $||C-C'||_2 > \varepsilon$}{
    $C' \leftarrow C$\;
    \tcp*[l]{Always pick the best initial direction}
    \eIf{ $L_{r}<L_{c}$ }{
      $C \leftarrow$ {\sc refinement}($A, C, p$) \;
    }{
      $C \leftarrow$ {\sc refinement}($A^T, C, p$) \;
    }
    $i \leftarrow i+1$ \;
  }

  \Return $C$
  \caption{{\sc PBD}($A, p$)}
  \label{alg:pbd}
\end{small}
\end{algorithm2e}

\subsubsection{Pick best (in each) iteration (PBI)}
\label{subsec:pbi}

algorithm applies refinement on both row-based and column partitions and always
chooses the best partition vector for the next iteration.
If that partition vector improves the current best solution, PBI algorithm
updates the partition vector that stores the best solution achieved.
This procedure is presented in Algorithm~\ref{alg:pbi}.

Tracking the load imbalance creates an opportunity to output a
better partitioning. However, tracking comes up with a computational cost ($O(m)$).
Hence, PBI algorithm is more expensive than PBD algorithm.

\begin{algorithm2e}[ht]
\begin{small}
  \tcp*[l]{Initialize partition vectors}
  $C_c[0] = 0$; $C_c[j] = n$, for $ 1 \leq j \leq p+1$ \tcp*[r]{Col. based}
  $C_r[0] = 0$; $C_r[j] = n$, for $ 1 \leq j \leq p+1$ \tcp*[r]{Row based}
  $C_b[0] = 0$; $C_b[j] = n$, for $ 1 \leq j \leq p+1$ \tcp*[r]{Best}

  $i \leftarrow 0$ \;

  \While{ $i<\tau$ }{
    \tcp*[l]{Compute row based and column based partition vectors.}
    $C_r \leftarrow$ {\sc refinement}($A, C_c, p$) \;
    $C_c \leftarrow$ {\sc refinement}($A^T, C_r, p$) \;

    \tcp*[l]{Pick the best partition vector}
    \eIf{$\lambda(A, C_c, C_c ) < \lambda(A, C_r, C_r ) $}{
      $C_r \leftarrow C_c$ \;
    }{
      $C_c \leftarrow C_r$ \;
    }

    \tcp*[l]{Update the best partition vector, if improved}
    \If{$\lambda(A, C_c, C_c) < \lambda(A, C_b, C_b)$}{
      $C_b \leftarrow C_c$ \;
    }

    $i \leftarrow i+1$ \;
  }

  \Return $C_{b}$
  \caption{{\sc PBI}($A, p$)}
  \label{alg:pbi}
\end{small}
\end{algorithm2e}

\subsubsection{Probe target cut (PTC)}
\label{subsec:ptc}

probes for the largest possible cut point in each step using a
two-dimensional probe algorithm. In refinement based algorithms (PBI and PBD)
mapping the problem from two-dimensional space to one-dimension
ends up with losing information. PTC algorithm tries to overcome this
problem by seeking better cut points in two-dimensional space. In high-level view
PTC algorithm is inspired from Nicol's~\cite{Nicol94-JPDC} probe-based
one-dimensional partitioning algorithm however PTC algorithm operates on
two-dimensional space on diagonal direction in a greedy manner and due to
non-convex structure of the problem it doesn't guarantee an optimal solution.
Algorithm~\ref{alg:probe} presents the
two-dimensional probe algorithm. The elements of $C$ are found
through binary search, $\beta$, on the matrix.
In this algorithm, $\beta(A, C, i, \ell)$, searches $A$ in the range $[C[i-1],n]$
to compute the largest cut point, $C[i]=j$ such that $\lambda(A, C, i) \leq \ell$ and for
$C[i]=j+1$, $\lambda(A,C,i)>\ell$. Algorithm~\ref{alg:probe} returns {\tt true} if at the end
partition vector has $p$ intervals and the load imbalance is less than $\ell$.
In each step, $i$, PTC algorithm seeks for the largest cut point in the range
$[C[i-1],n]$ for which Probe Algorithm~\ref{alg:probe} returns {\tt true}. PTC stores the
load imbalances of these steps in an array, $B$. After $p-1$ steps, PTC algorithm
computes the minimum of $B$ and constructs partition vector $C$ using a binary
search-based approach similar to Algorithm~\ref{alg:probe}.
This procedure is presented in Algorithm~\ref{alg:ptc}.

\begin{algorithm2e}[ht]
\begin{small}
  \tcp*[l]{Initialize partition vector}
  $C[0] = 0$; $C[p] = n$ \;

  \For{$i=1$ {\bf to} $p-1$}{
    $C[i] \leftarrow \beta(A, C, i, \ell)$\;
  }

  \eIf{$\lambda(A, C, C) \leq \ell$}{
    \Return {\tt true}
  }{
    \Return {\tt false}
  }

  \caption{{\sc probe}($A, p, \ell$)}
  \label{alg:probe}
\end{small}
\end{algorithm2e}

Note that, probe algorithm can return {\tt false} even if there
exists a valid partitioning for a given target load imbalance $\ell$. Because PTC algorithm
always searches for the largest cut point and that may cause missing the optimal solution.
Hence, this algorithm can also be stuck in a local optimum. However,
PTC algorithm considers more cases in a two-dimensional fashion. Therefore,
PTC is expected to produce a better partitioning than PBD and PBI algorithms.
The major disadvantage of this algorithm is its computational complexity.

\begin{algorithm2e}[ht]
\begin{small}
  \tcp*[l]{Initialize temporary partition vector}
  $C[0] = 0$; $C[j] = n$, for $ 1 \leq j \leq p+1$ \;

  \tcp*[l]{An array to store load imbalances}
  $B[j] = 0$, for $ 0 \leq j < p$ \;

  \For{$i=1$ {\bf to} $p-1$}{
    $l \leftarrow C[i-1]$ \;
    $r \leftarrow n$ \;
    \tcp*[l]{Probe in binary search fashion}
    \While{$l<r$}{
      $m \leftarrow (l+r)/2$ \;
      $C[i] \leftarrow m$ \;
      $\ell \leftarrow \lambda(A, C, C, i)$ \;
      \eIf{{\sc probe($A, p, \ell$)}}{
        $r \leftarrow m$ \;
        $B[i-1] \leftarrow \ell$ \;
      }{
        $l \leftarrow m+1$ \;
      }
    }
    $C[i] \leftarrow r$ \;
  }

  \tcp*[l]{Find the minimum load imbalance}
  $B_{min} \leftarrow \min_{j}\{B[j]\}$\;

  $C[0] = 0$; $C[p] = n$ \;

  \tcp*[l]{Construct partition vector}
  \For{$i=1$ {\bf to} $p$}{
    $C[i] \leftarrow \beta(A, C, i, B_{min})$\;
  }

  \Return $C$
  \caption{{\sc PTC}($A, p$)}
  \label{alg:ptc}
\end{small}
\end{algorithm2e}

\subsection{Algorithms for the {\sc minNumCuts} problem}

Given a matrix, $A$, and an integer, $Z$, \mNC problem aims to output
a partition vector, $C$, with the minimum number of intervals, $p$, where the maximum
load of a tile in the corresponding partitioning is less then $Z$, i.e.,
$\max_{0 \leq i,j \leq p}\{T_{i,j}\} \leq Z$.
We propose two algorithms to solve this problem. These algorithms are
variations of the ones that we propose for \mLI problem.

\subsubsection{Bound target load (BTL)}
\label{subsec:btl}

algorithm, displayed in Algorithm~\ref{alg:btl}, searches for the minimal $p$
intervals using a binary search-based approach (Algorithm~\ref{alg:bin}).
In the worst-case, the maximum loaded tile may be filled
with nonzeros. Therefore, the initial upper bound for the search
space can be defined as $p\in[1,\frac{n}{\sqrt{Z}}]$.
BTL algorithm first reduces this upper bound by using cheap uniform
partitioning.
Then using this reduced search space, BTL algorithm searches for the minimal $p$
intervals using PBD (or PBI) algorithm in the same fashion.

\begin{algorithm2e}[ht]
\begin{small}
  \tcp*[l]{$u$ is the upper bound for number of intervals}
  \tcp*[l]{Initial upper bound: Tile is fully connected}
  $u \leftarrow \frac{n}{\sqrt{Z}}$ \;

  \tcp*[l]{Decreasing upper bound using a cheap algorithm}
  $u \leftarrow $ {\sc findUpperBound}($A, Z, 1, u,$ {\sc UNI})\;

  \tcp*[l]{Using PBD (PBI) search for a lower upper bound}
  $u \leftarrow $ {\sc findUpperBound}($A, Z, 1, u,$ {\sc PBD})\;

  \Return {\sc PBD}($A, u$)
  \caption{{\sc BTL}($A, Z$)}
  \label{alg:btl}
\end{small}
\end{algorithm2e}

\begin{algorithm2e}[ht]
\begin{small}

  \While{$l<r$}{
    $p \leftarrow (l+r)/2$ \;
    $C \leftarrow  f(A, p)$ \;
    $\ell \leftarrow \frac{Z \times p^2}{m}$ \;
    $\ell' \leftarrow \lambda(A, C, C)$ \;
    \eIf{$\ell'<\ell$}{
      $r \leftarrow p$ \;
    }{
      $l \leftarrow p+1$ \;
    }
  }

  \Return $r$
  \caption{{\sc findUpperBound}($A, Z, l, r, f(\dot)$)}
  \label{alg:bin}
\end{small}
\end{algorithm2e}

\subsubsection{Probe target load (PTL)}
\label{subsec:ptl}

algorithm slightly modifies two-dimensional probe algorithm. In each step, $k$, PTL
algorithm searches for the largest cut point, $C[k] = l$ in the range $[C[k-1], n]$
that assures all of the tiles until that cut point have less than $Z$
number of nonzeros, $\forall_{i,j\leq k} |T_{i,j}|\leq Z$.
This procedure is described in Algorithm~\ref{alg:ptl}.

\begin{algorithm2e}[ht]
\begin{small}
  \tcp*[l]{Initially we don't know partition vector's size}
  $C[0] = 0$ \;
  $i \leftarrow 1$\;
  \While{$C[i-1]\neq n$}{
    $C[i] \leftarrow  \beta(A, C, i, Z) $\;
    $i \leftarrow i+1$ \;
  }

  \Return $C$

  \caption{{\sc PTL}($A, Z$)}
  \label{alg:ptl}
\end{small}
\end{algorithm2e}

\subsection{Complexity Analysis.}
\label{subsec:complexity}

\Cref{tab:complexity} displays the computational complexity of the algorithms we have
used in this work.
Each iteration of the iterative refinement algorithm~\cite{Manne96-IWAPC,Nicol94-JPDC}
(Algorithm~\ref{alg:rowparti}) has a worst-case complexity of
$O(q(p\log \frac{n}{p})^2 + p(q\log \frac{ n}{q})^2)$~\cite{Saule12-JPDC-spart} for
unsymmetric rectilinear partitioning. The Algorithm is guaranteed to converge with
at most $n^2$ iterations. However, as noted in these earlier work,
in our experiments we observed that algorithm converges very quickly, and hence
for the sake of fairness we have decided to use the same limit on the number of iterations, $\tau$.
For the symmetric case, where $p=q$, refinement algorithm runs in
$O(p^3(\log \frac{n}{p})^2)$. and this what we displayed in \Cref{tab:complexity}.

PBD algorithm first runs Algorithm~\ref{alg:rowparti} and then computes
the load imbalance. These operations can be computed in
$O(p^3(\log \frac{n}{p})^2)$ and in $O(m)$ respectively. In the worst-case,
Algorithm~\ref{alg:rowparti} is called $\tau$ times. Hence, PBD algorithm runs in
$O(m + \tau p^3(\log \frac{n}{p})^2)$.

PBI algorithm runs Algorithm~\ref{alg:rowparti} and then computes
the load imbalance in each of the $\tau$ iterations.
Hence, PBI algorithm runs in
$O( \tau(m+p^3(\log \frac{n}{p})^2) )$.

Probe algorithm (Algorithm~\ref{alg:probe}) does $O(m \log n)$ computations in
the worst-case to find a cut point. Since there are $O(p)$ cut points
probe algorithm runs in
$O(p m\log n)$. PTC algorithm (Algorithm~\ref{alg:ptc}) calls Probe algorithm
at most $\log n$ times for each $i \in [1,p-1]$. Hence, PTC algorithm runs
in $O(m(p\log n)^2)$.

BTL algorithm initially tries to reduce search space using cheap uniform
partitioning, however in the worst-case uniform partitioning may not be able
to reduce search space. Hence, in the worst-case BTL algorithm runs
$\log \frac{n}{\sqrt{Z}}$ times uniform and PBD (or PBI) algorithm.
In the worst-case $p=\frac{n}{\sqrt{Z}}$  (i.e., when there is fully dense tile).
So, BTL algorithm runs in
$O( \log \frac{n}{\sqrt{Z}} (m + \tau\frac{n}{\sqrt{Z}}^3(\log \sqrt{Z})^2) )$
when PBD is used as the secondary algorithm and in
$O( \log \frac{n}{\sqrt{Z}} \tau ( m+ \frac{n}{\sqrt{Z}}^3( \log \sqrt{Z})^2) )$
when PBI is used as the secondary algorithm.

PTL algorithm has the same complexity with Algorithm~\ref{alg:probe} and in the
worst-case $p=\frac{n}{\sqrt{Z}}$.
Hence, PTL algorithm runs in; $O( \frac{n}{\sqrt{Z}} m\log n )$.

In addition to our proposed algorithms, we use two reference partitioning
algorithms as baselines. First,
{NIC} refers to Nicol's rectilinear partitioning algorithm~\cite{Nicol94-JPDC}.
This algorithm outputs a partition vector for each dimension, hence it does not
output symmetric partitioning. The complexity of this algorithm is provided
in~\cite{Nicol94-JPDC} and listed in \Cref{tab:complexity}.
Second, {UNI} refers to uniform partitioning. This is the simplest
checker board partitioning, where each tile has an equal number of
rows and columns. UNI algorithm runs in constant time.

\begin{table}[!ht]
\begin{small}
\begin{tabular}{l l}
\hline
\textbf{Algorithm} & \textbf{Worst-case complexity} \\
\hline
\bf NIC & $O( \tau p^3(\log \frac{n}{p})^2)$ \\
\bf PBD & $O( m + \tau p^3(\log \frac{n}{p})^2 )$ \\
\bf PBI & $O( \tau(m+p^3(\log \frac{n}{p})^2) )$ \\
\bf PTC & $O( m(p\log n)^2 )$ \\
\bf BTL+PBD & $O( \log \frac{n}{\sqrt{Z}} (m + \tau \frac{n}{\sqrt{Z}}^3(\log \sqrt{Z})^2) )$ \\
\bf BTL+PBI & $O( \log \frac{n}{\sqrt{Z}} \tau ( m+ \frac{n}{\sqrt{Z}}^3(\log \sqrt{Z})^2) )$ \\
\bf PTL & $O( \frac{n}{\sqrt{Z}} m\log n )$ \\
\hline
\end{tabular}
\end{small}
\caption{Complexity of the Algorithms.}
\label{tab:complexity}
\end{table}

\setlength{\tabcolsep}{3pt}
\begin{table*}[!ht]
  \supersmallfont
    \begin{center}
      \begin{tabular}{ | l|| 
          r r ||
          c c c c c ||
          c c c c c ||
          c c c c c |
        }
        \hline

        \multirow{2}{*}{\textbf{Data Set}}  &
        \multirow{2}{*}{$\boldsymbol{n}$}   &
        \multirow{2}{*}{$\boldsymbol{ m}$}   &
        \multicolumn{5}{c||}{\textbf{Natural Order}}  &
        \multicolumn{5}{c||}{\textbf{Degree Order}}  &
        \multicolumn{5}{c|}{\textbf{RCM Order}}  \\

        & &
        & NIC & UNI & PBD & PBI & PTC
        & NIC & UNI & PBD & PBI & PTC
        & NIC & UNI & PBD & PBI & PTC\\ \hline \hline

    cit-HepTh  &  27,770  &  352,285
    & 1.7 & \cellcolor{yellow!50} 1.1 & 1.4 & 1.7 & 0.6
    & 1.2 & 8.5 & \cellcolor{yellow!50} 0.8 & 1.2 & \cellcolor{green!50} 0.5
    & 1.7 & 3.8 & \cellcolor{yellow!50} 1.5 & 1.7 & 0.6 \\ \hline

    email-EuAll  &  265,214  &  364,481
    & 2.2 & \cellcolor{yellow!50} 1.1 & 1.9 & 2.2 & \cellcolor{green!50} 0.2
    & 3.5 & 13.0 & \cellcolor{yellow!50} 3.5 & 3.5 & 3.3
    & 3.3 & 9.1 & 3.3 & \cellcolor{yellow!50} 3.3 & 2.4 \\ \hline

    soc-Epinions1  &  75,879  &  405,740
    & 1.7 & 2.1 & \cellcolor{yellow!50} 1.3 & 1.7 & 0.7
    & 1.2 & 23.0 & \cellcolor{yellow!50} 0.7 & 1.2 & \cellcolor{green!50} 0.4
    & 2.0 & 16.6 & \cellcolor{yellow!50} 1.7 & 2.0 & 0.7 \\ \hline

    cit-HepPh  &  34,546  &  420,877
    & 1.7 & \cellcolor{yellow!50} 0.8 & 1.3 & 1.7 & 0.6
    & 1.3 & 6.4 & \cellcolor{yellow!50} 0.7 & 1.3 & \cellcolor{green!50} 0.3
    & 1.8 & 3.6 & \cellcolor{yellow!50} 1.7 & 1.8 & 0.7 \\ \hline

    soc-Slashdot0811  &  77,360  &  469,180
    & 1.8 & 2.4 & \cellcolor{yellow!50} 1.3 & 1.8 & 0.7
    & 1.7 & 18.6 & \cellcolor{yellow!50} 0.8 & 1.6 & \cellcolor{green!50} 0.6
    & 1.9 & 12.6 & \cellcolor{yellow!50} 1.5 & 1.9 & 0.6 \\ \hline

    soc-Slashdot0902  &  82,168  &  504,230
    & 1.8 & 2.6 & \cellcolor{yellow!50} 1.3 & 1.8 & 0.7
    & 1.6 & 18.5 & \cellcolor{yellow!50} 0.7 & 1.6 & 0.4
    & 1.8 & 12.4 & \cellcolor{yellow!50} 1.4 & 1.8 & \cellcolor{green!50} 0.3 \\ \hline

    flickrEdges  &  105,938  &  2,316,948
    & 1.6 & 1.2 & \cellcolor{yellow!50} 1.1 & 1.6 & \cellcolor{green!50} 0.6
    & 1.9 & 25.2 & \cellcolor{yellow!50} 1.8 & 1.9 & 1.0
    & \cellcolor{yellow!50} 2.1 & 21.2 & 2.3 & 2.3 & 1.8 \\ \hline

    amazon0312  &  400,727  &  2,349,869
    & 1.8 & \cellcolor{yellow!50} 1.1 & 1.6 & 1.8 & \cellcolor{green!50} 0.4
    & 1.8 & 2.5 & \cellcolor{yellow!50} 1.2 & 1.8 & 0.7
    & 2.0 & 2.0 & \cellcolor{yellow!50} 1.9 & 2.0 & 0.8 \\ \hline

    amazon0505  &  410,236  &  2,439,437
    & 1.8 & \cellcolor{yellow!50} 1.2 & 1.6 & 1.8 & \cellcolor{green!50} 0.4
    & 1.8 & 2.5 & \cellcolor{yellow!50} 1.2 & 1.8 & 0.6
    & 2.0 & 1.9 & \cellcolor{yellow!50} 1.9 & 2.0 & 0.7 \\ \hline

    amazon0601  &  403,394  &  2,443,408
    & 2.1 & 3.0 & \cellcolor{yellow!50} 1.9 & 2.1 & 0.6
    & 1.9 & 2.4 & \cellcolor{yellow!50} 1.2 & 1.9 & \cellcolor{green!50} 0.6
    & 2.0 & 2.0 & \cellcolor{yellow!50} 1.9 & 2.0 & 0.8 \\ \hline

    scale18  &  174,147  &  3,800,348
    & 1.7 & 1.4 & \cellcolor{yellow!50} 0.9 & 1.7 & \cellcolor{green!50} 0.8
    & 1.8 & 22.2 & \cellcolor{yellow!50} 1.1 & 1.8 & 1.1
    & 1.9 & 17.2 & \cellcolor{yellow!50} 1.0 & 1.9 & 0.7 \\ \hline

    scale19  &  335,318  &  7,729,675
    & 2.2 & 6.7 & 2.2 & \cellcolor{yellow!50} 2.2 & 1.1
    & 2.4 & 13.5 & \cellcolor{yellow!50} 2.0 & 2.4 & 1.0
    & 2.2 & 9.9 & \cellcolor{yellow!50} 2.0 & 2.2 & \cellcolor{green!50} 0.6 \\ \hline

    as-Skitter  &  1,696,415  &  11,095,298
    & 1.7 & 1.6 & \cellcolor{yellow!50} 1.0 & 1.7 & \cellcolor{green!50} 0.9
    & 1.8 & 23.7 & \cellcolor{yellow!50} 1.0 & 1.8 & 1.0
    & 1.9 & 18.5 & \cellcolor{yellow!50} 1.3 & 1.9 & 1.2 \\ \hline

    scale20  &  645,820  &  15,680,861
    & 1.7 & 1.4 & \cellcolor{yellow!50} 1.0 & 1.7 & \cellcolor{green!50} 0.9
    & 1.8 & 24.1 & \cellcolor{yellow!50} 1.0 & 1.8 & 1.0
    & 1.9 & 19.4 & \cellcolor{yellow!50} 1.3 & 1.9 & 1.2 \\ \hline

    cit-Patents  &  3,774,768  &  16,518,947
    & 1.9 & 2.2 & \cellcolor{yellow!50} 1.8 & 1.9 & 1.0
    & 1.6 & 6.7 & \cellcolor{yellow!50} 1.2 & 1.6 & \cellcolor{green!50} 0.7
    & 2.0 & 3.3 & \cellcolor{yellow!50} 1.9 & 2.0 & 0.7 \\ \hline

    scale21  &  1,243,072  &  31,731,650
    & 1.7 & 1.3 & \cellcolor{yellow!50} 1.0 & 1.7 & 0.9
    & 1.8 & 24.8 & \cellcolor{yellow!50} 1.0 & 1.8 & 1.0
    & 1.9 & 20.4 & \cellcolor{yellow!50} 1.0 & 1.9 & \cellcolor{green!50} 0.8 \\ \hline

    soc-LiveJournal1  &  4,847,571  &  42,851,237
    & 1.9 & 7.7 & \cellcolor{yellow!50} 1.7 & 1.9 & \cellcolor{green!50} 0.4
    & 1.3 & 15.6 & \cellcolor{yellow!50} 1.0 & 1.3 & 0.6
    & 1.8 & 7.2 & \cellcolor{yellow!50} 1.6 & 1.8 & 0.5 \\ \hline

    wb-edu  &  9,845,725 &  46,236,105
    & 3.4 & 4.0 & 3.4 & \cellcolor{yellow!50} 3.4 & 3.2
    & 2.6 & 13.2 & \cellcolor{yellow!50} 2.6 & 2.6 & \cellcolor{green!50} 0.5
    & 3.0 & 4.0 & \cellcolor{yellow!50} 3.0 & 3.0 & 1.7 \\ \hline

    twitter  &  61,578,414  &  1,202,513,046
    & 1.7 & 2.0 & \cellcolor{yellow!50} 1.2 & 1.7 & \cellcolor{green!50} 0.3
    & 2.3 & 23.1 & \cellcolor{yellow!50} 2.1 & 2.3 & 1.7
    & 1.5 & 5.2 & \cellcolor{yellow!50} 1.4 & 1.5 & 1.0 \\ \hline

    friendster  &  65,608,366  &  1,806,067,135
    & 1.3 & 2.7 & \cellcolor{yellow!50} 1.1 & 1.3 & 0.6
    & 1.2 & 17.5 & \cellcolor{yellow!50} 0.6 & 1.2 & \cellcolor{green!50} 0.4
    & 1.3 & 10.6 & \cellcolor{yellow!50} 1.0 & 1.3 & 0.5 \\ \hline \hline

    \multicolumn{3}{|r}{\bf Geomean:}
    & {\bf 1.8} & {\bf 1.9} & {\bf 1.4} & {\bf 1.8} & {\bf 0.7}
    & {\bf 1.8} & {\bf 12.2} & {\bf 1.2} & {\bf 1.8} & {\bf 0.7}
    & {\bf 2.0} & {\bf 7.5} & {\bf 1.6} & {\bf 2.0} & {\bf 0.8} \\ \hline

      \end{tabular}
    \caption{Properties of the dataset and load imbalance for $8 \times 8$
    partitioning on three different layouts and five different algorithms.
    NIC - Nicol's rectilinear partitioning. UNI - Uniform partitioning.
    Green - Best load imbalance for each graph.
    Yellow  - Best load imbalance among four algorithms for each graph and vertex ordering.}
    \label{table:dataset}

        \end{center}
\end{table*}

\section{Experimental Evaluation}

The partitioning algorithms presented for \mLI and \mNC problems are implemented
in the C++ programming language and compiled with GCC version 7.2. The experiments
are conducted on a server that has four, 14-core Intel Xeon E7-4850 2.20GHz
processors, 2TB of memory, 1TB disk space, running Ubuntu GNU/Linux with
kernel 4.8.0.

We have performed an extensive evaluation of the proposed algorithms on 16
different real-world and 4 different synthetic (RMAT) graphs coming
from SuiteSparse Matrix Collection (formerly known as UFL)~\cite{Davis11-TOMS},
SNAP~\footnote{SNAP Datasets: \url{http://snap.stanford.edu/data}}, and
DARPA Graph Challange~\footnote{GraphChallenge Datasets: \url{https://graphchallenge.mit.edu/data-sets}}.
Properties of these datasets, along with the load balances found
by different algorithms on different layouts of these matrices are presented in \Cref{table:dataset}.
In the experiments, we used $\tau=20$ and $\epsilon=0.0001$.

\begin{figure}[!ht]
  \centering
  \subfigure[Natural Order]{\includegraphics[width=.31\linewidth]{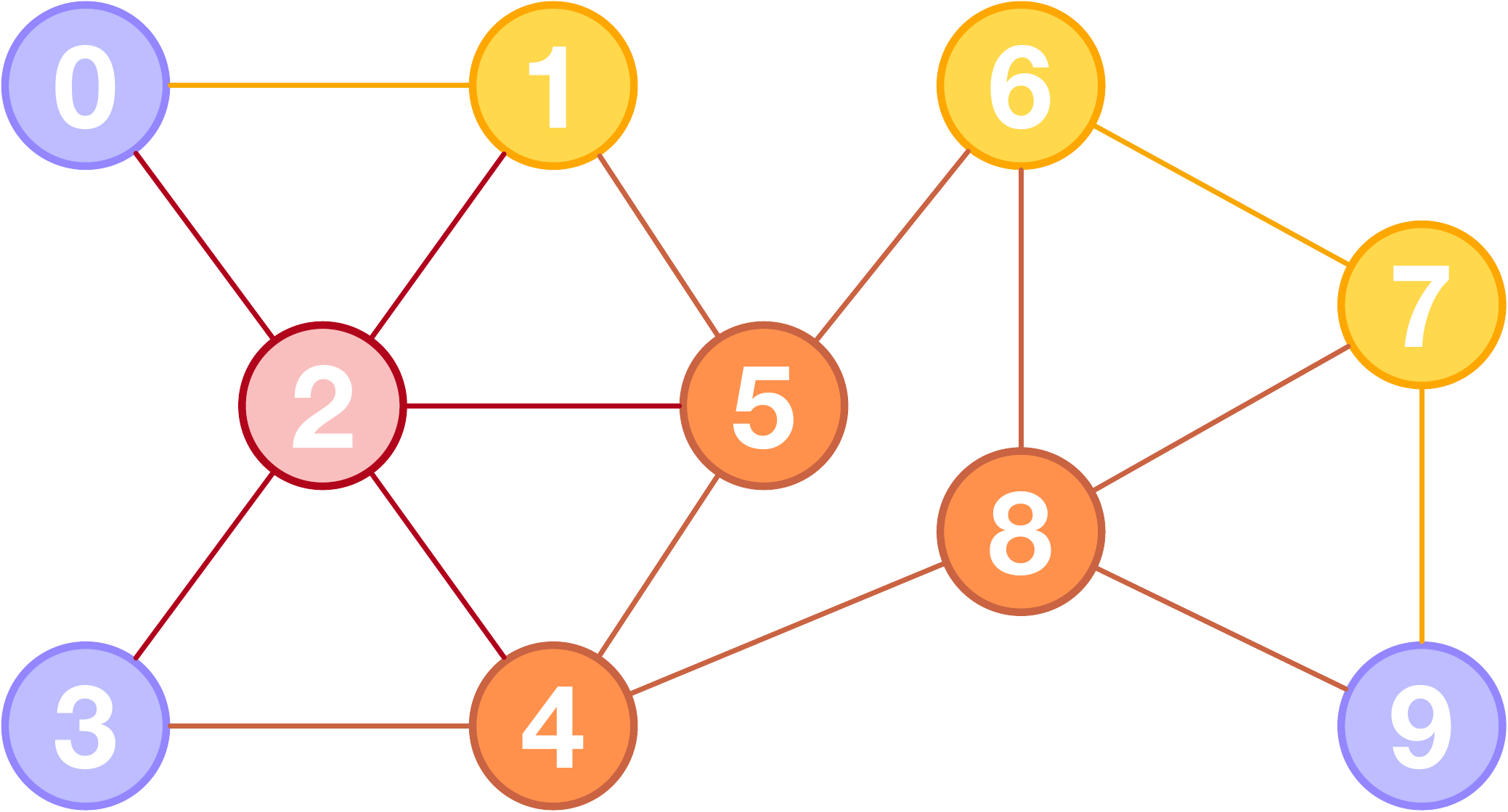}
  \label{fig:toy-nat}}
  \subfigure[Degree Order]{\includegraphics[width=.31\linewidth]{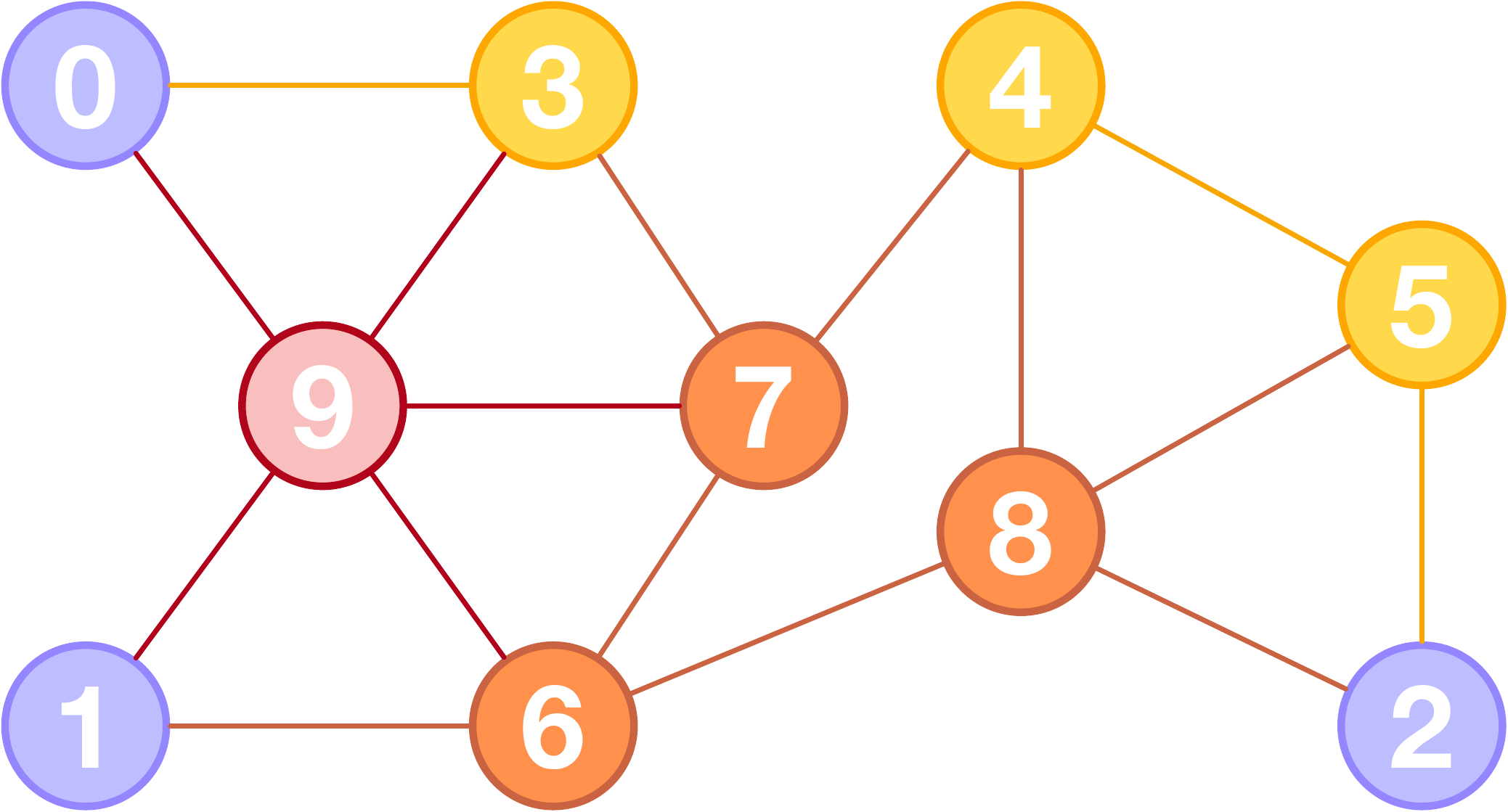}
  \label{fig:toy-deg}}
  \subfigure[RCM Order]{\includegraphics[width=.31\linewidth]{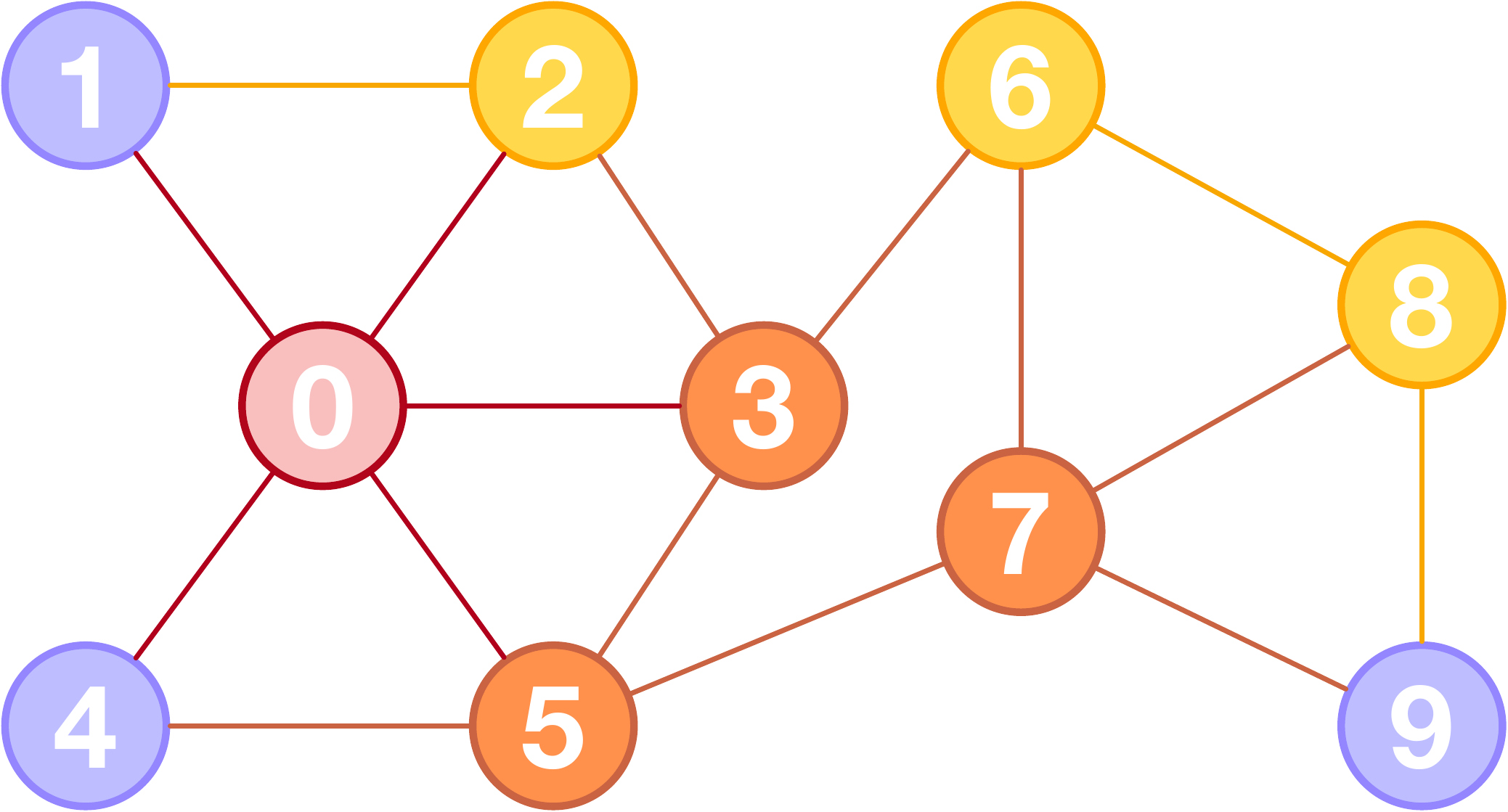}
  \label{fig:toy-rcm}}

  \subfigure[Matrix Form]{\includegraphics[width=.28\linewidth]{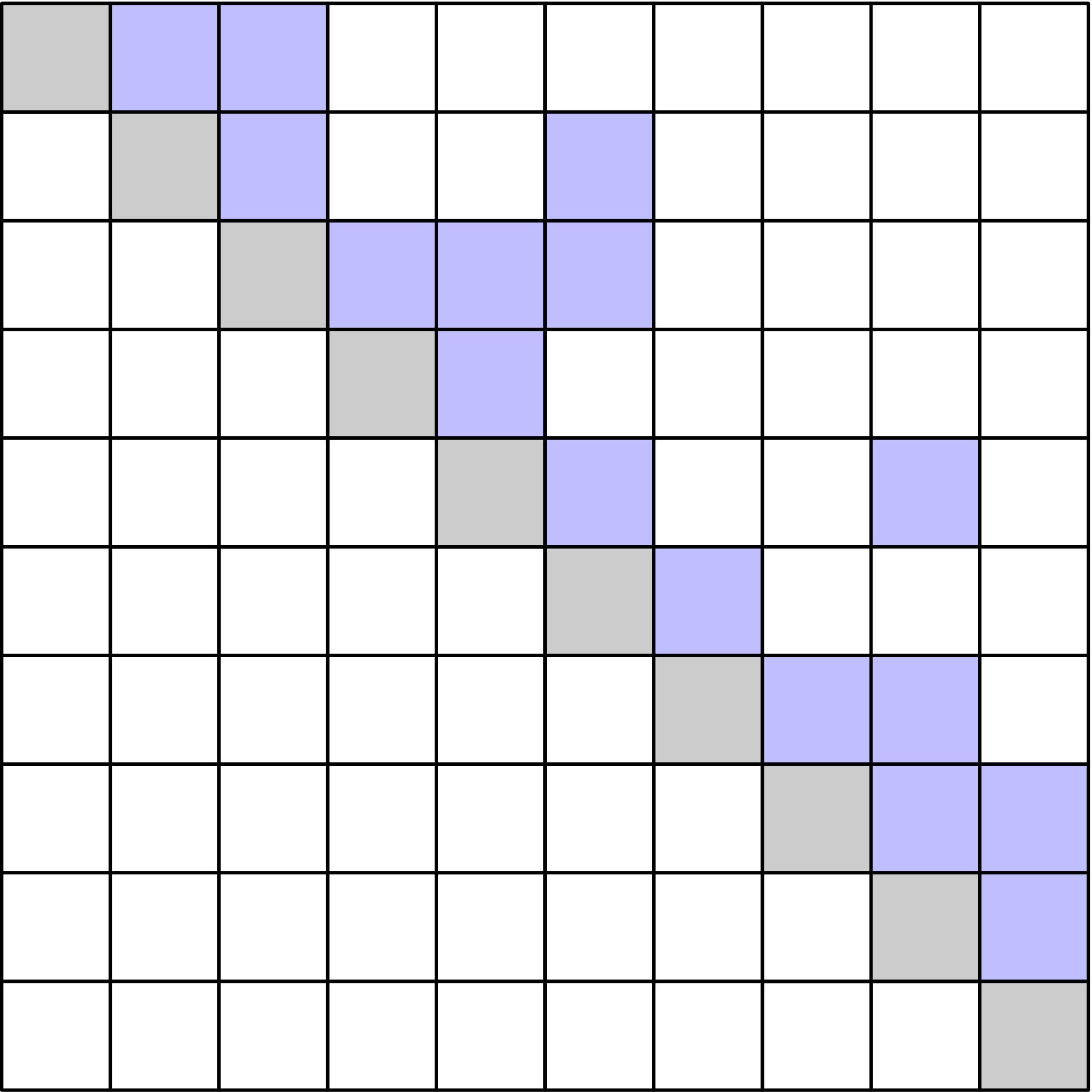}
  \label{fig:toy-nat-mtx}} \hspace{.5em}
  \subfigure[Matrix Form]{\includegraphics[width=.28\linewidth]{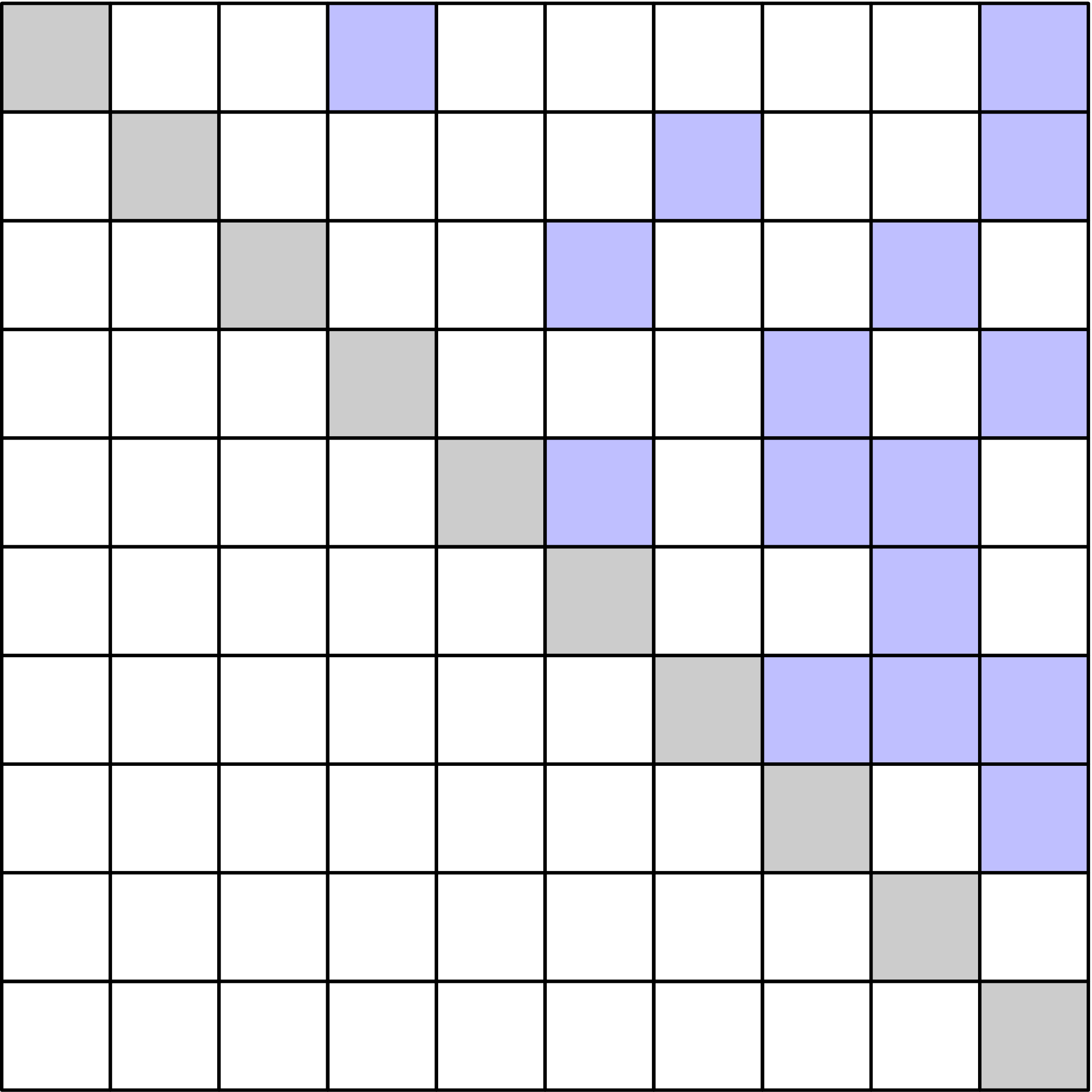}
  \label{fig:toy-deg-mtx}} \hspace{.5em}
  \subfigure[Matrix Form]{\includegraphics[width=.28\linewidth]{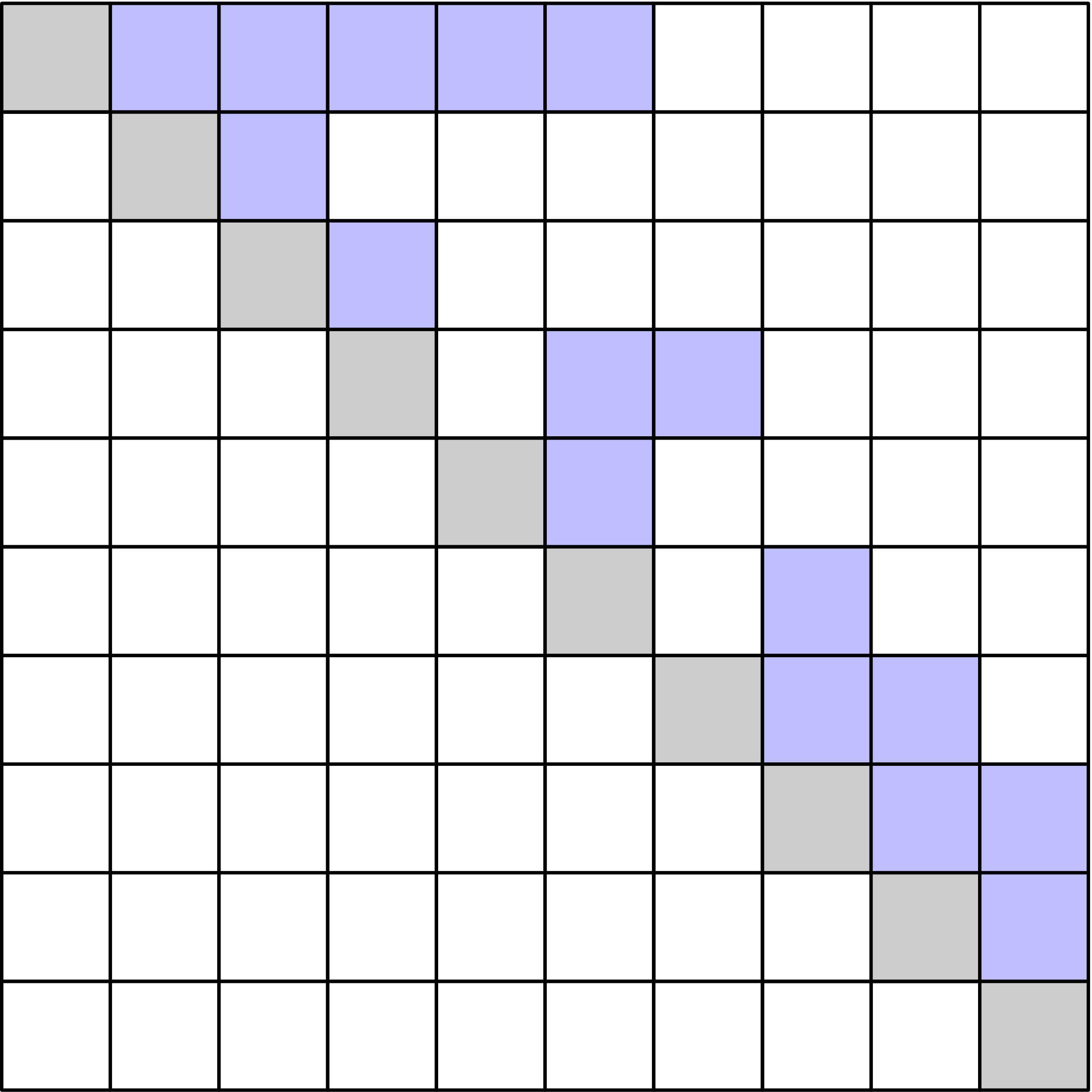}
  \label{fig:toy-rcm-mtx}}

  \caption{Three different vertex orderings (upper) and their adjacency matrix representations (lower)}
  \label{fig:exorder}
\end{figure}

In our experiments, we included three different vertex ordering techniques
before giving the adjacency matrix to respective partitioner:
the natural order of the vertices (NAT), degree-based ordering of the vertices (DEG)
and Reverse Cuthill McKee (RCM) based ordering of the vertices.
\Cref{fig:exorder} illustrates these orderings along
with the adjacency matrix representations for a toy graph.

\subsection{Load imbalance evaluation.}

We first evaluate the proposed algorithms for the \mLI problem.
Table~\ref{table:dataset} reports load imbalances of
five different algorithms; NIC, UNI, PBD, PBI, and PTC,
on three different vertex ordering techniques, NAT, DEG and RCM for each graph.
In this experiment we chose $p=q=8$. Hence, every graph is partitioned into
$8 \times 8$ tiles.
In \Cref{table:dataset}, the best load imbalance for a graph
instance is highlighted using the green color.
As expected, PTC algorithm gives the best load imbalance in every graph instance.
The best load imbalance among with the other four algorithms; NIC, UNI, PBD,
and PBI, for each vertex ordering are highlighted using yellow color.
PBD algorithm gives the best performance among these four algorithms
in $54$ of the $60$ graph instances.
PBI and NIC algorithms give very similar load imbalances.

In \Cref{table:dataset}, the last row presents geometric means of the five
different algorithms on three different vertex orderings.
As shown in the table, the geometric means of PTC algorithm on NAT, DEG and RCM vertex
orderings are $0.7$, $0.7$ and $0.8$, respectively.
These results show that PTC algorithm is more resistant to the vertex order,
and hence, it can produce partitioning with similar qualities.
On the other hand, we observe a significant change in load imbalance (up to $40\%$)
for refinement based algorithms depending on the vertex order.
As expected, uniform partitioning performs poor on DEG vertex order due
to higher density in the bottom right portion of the adjacency matrix.

\subsection{Algorithm evaluation.}

\begin{figure*}[ht]
  \centering
  \subfigure[$2 \times 2$ Partitioning]{\includegraphics[width=.31\linewidth]{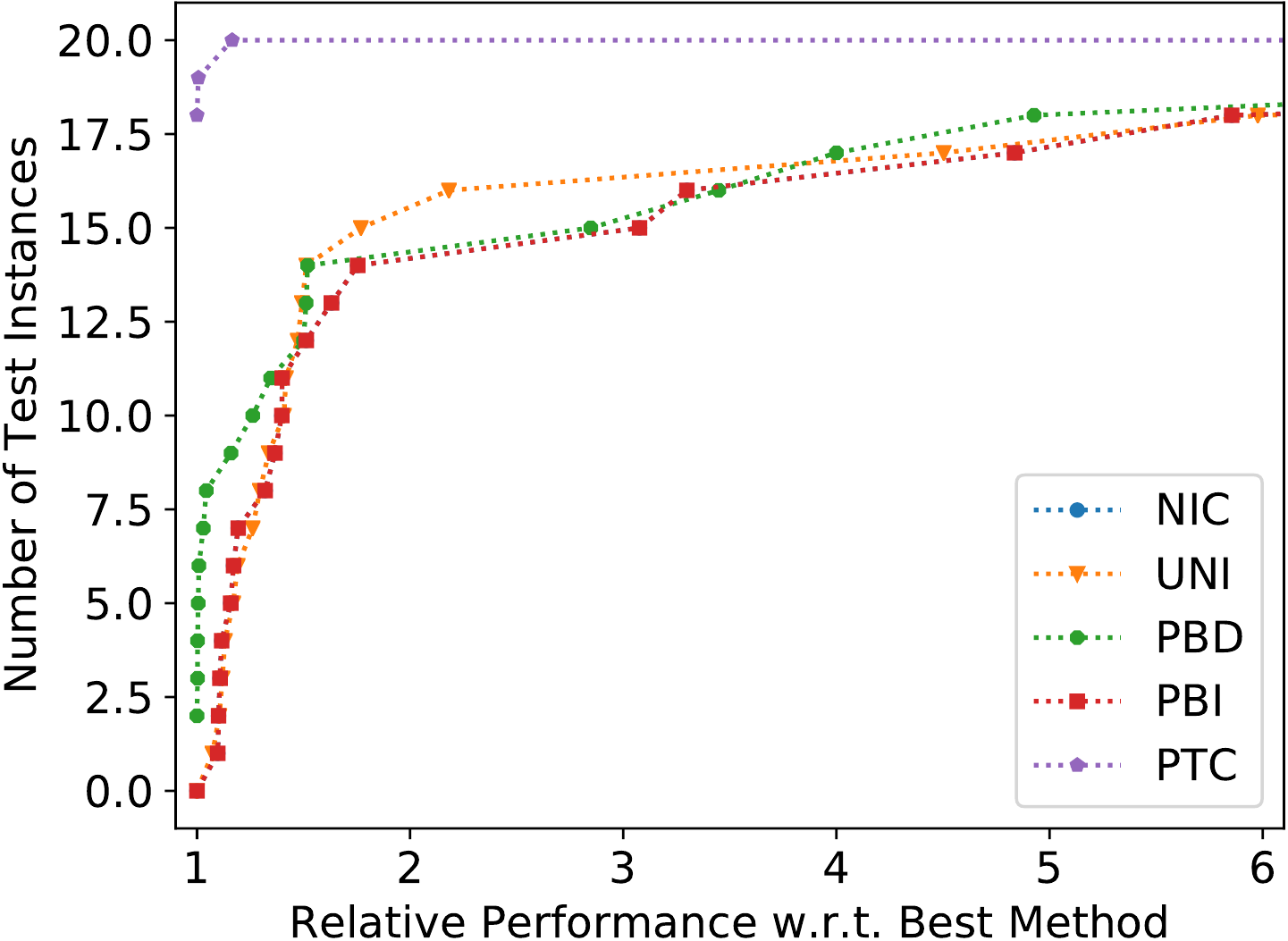}
  \label{fig:pp2-nat}}
  \subfigure[$4 \times 4$ Partitioning]{\includegraphics[width=.31\linewidth]{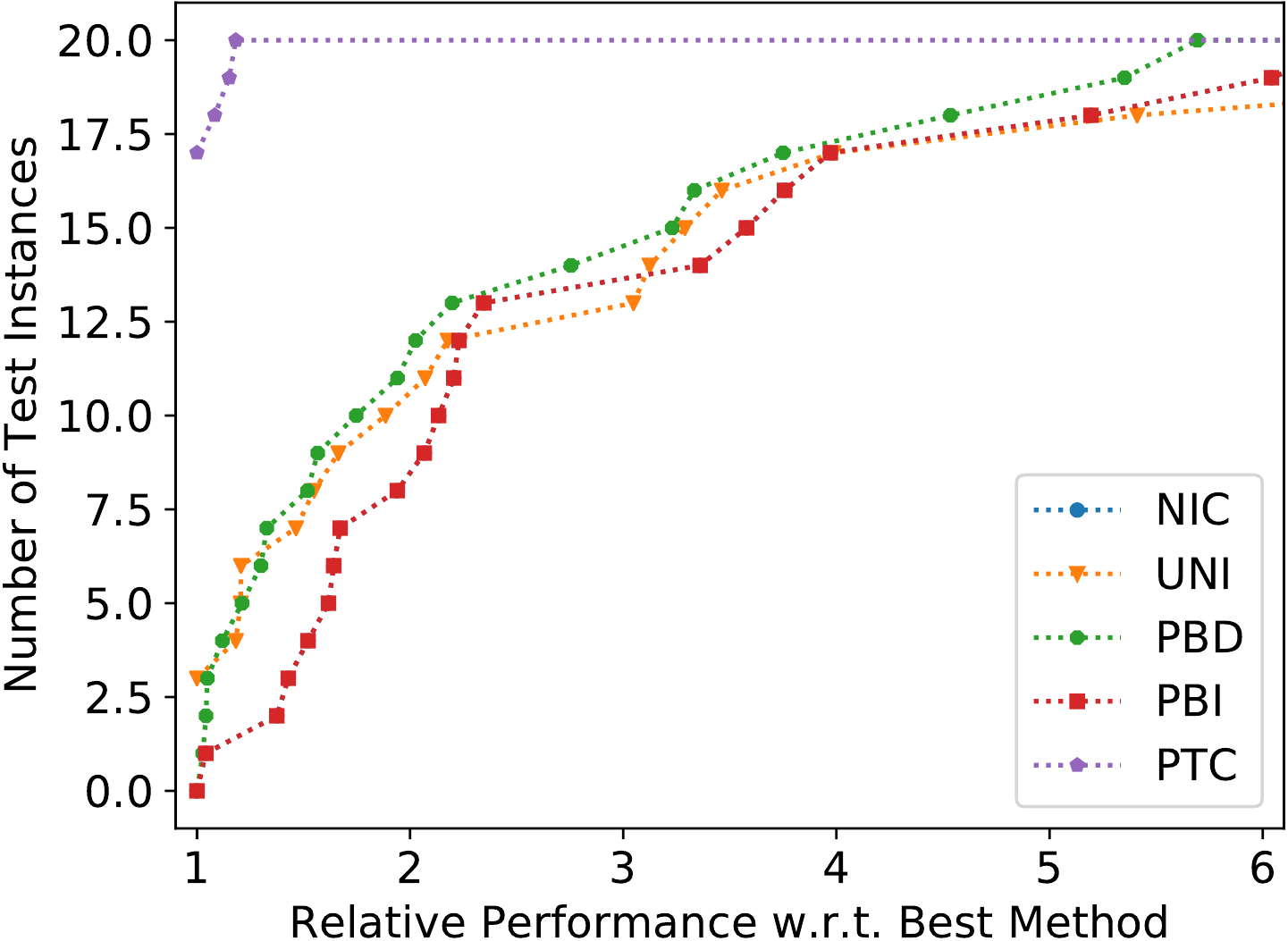}
  \label{fig:pp4-nat}}
  \subfigure[$8 \times 8$ Partitioning]{\includegraphics[width=.31\linewidth]{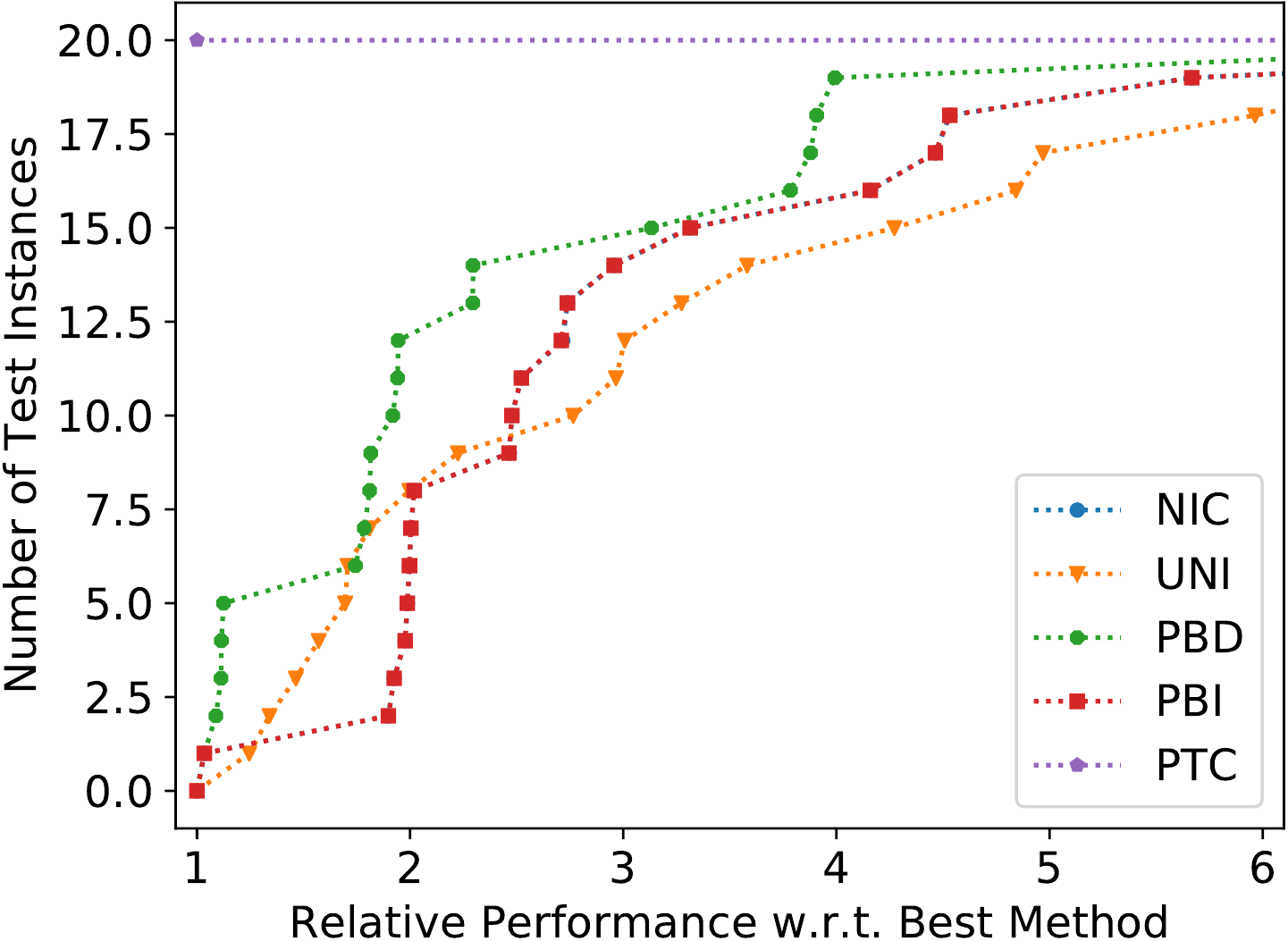}
  \label{fig:pp8-nat}}

  \caption{Performance profiles of UNI, NIC, PBD, PBI, PTC algorithms. NAT vertex order is used.
  \Cref{fig:pp2-nat} - $p=q=2$.
  \Cref{fig:pp4-nat} - $p=q=4$.
  \Cref{fig:pp8-nat} - $p=q=8$.}
  \label{fig:perfprof}
\end{figure*}

We evaluate relative load imbalance performances of NIC, UNI, PBD, PBI and PTC
algorithms. The aim is to illustrate the efficiency of the proposed algorithms with
respect to NIC and UNI. In this experiment, NAT vertex ordering is used and
we choose $p=\{2,4,8\}$.
\Cref{fig:perfprof} illustrates the performance profiles of the algorithms
for different $p$ values.
In the performance profiles, we plot the number of the test
instances (y-axis) in which an algorithm obtains a load imbalance on an instance that
is no larger than $x$ times (x-axis) the best load imbalance found by any algorithm
for that instance~\cite{Dolan02-MP}. Therefore, the higher a profile at a
given $x$ value, the better an algorithm is. We observe that in all cases
(\Cref{fig:pp2-nat}-\Cref{fig:pp8-nat}) PTC algorithm gives the best
performance in the majority of the test instances.
PBD algorithm becomes the second-best algorithm. We also observe
that NIC and PBI algorithms have almost identical profiles. Both of these
algorithms consider row-based and column-based directions in each iteration.
Therefore, probably they fall into the same local optimum and stuck.
As expected UNI algorithm performs the worst because of the irregularity of the graphs.

\subsection{Visualization of block distributions.}

\begin{figure*}[!ht]
  \centering
  \subfigure[NIC-LiveJournal]{\includegraphics[width=.23\linewidth]{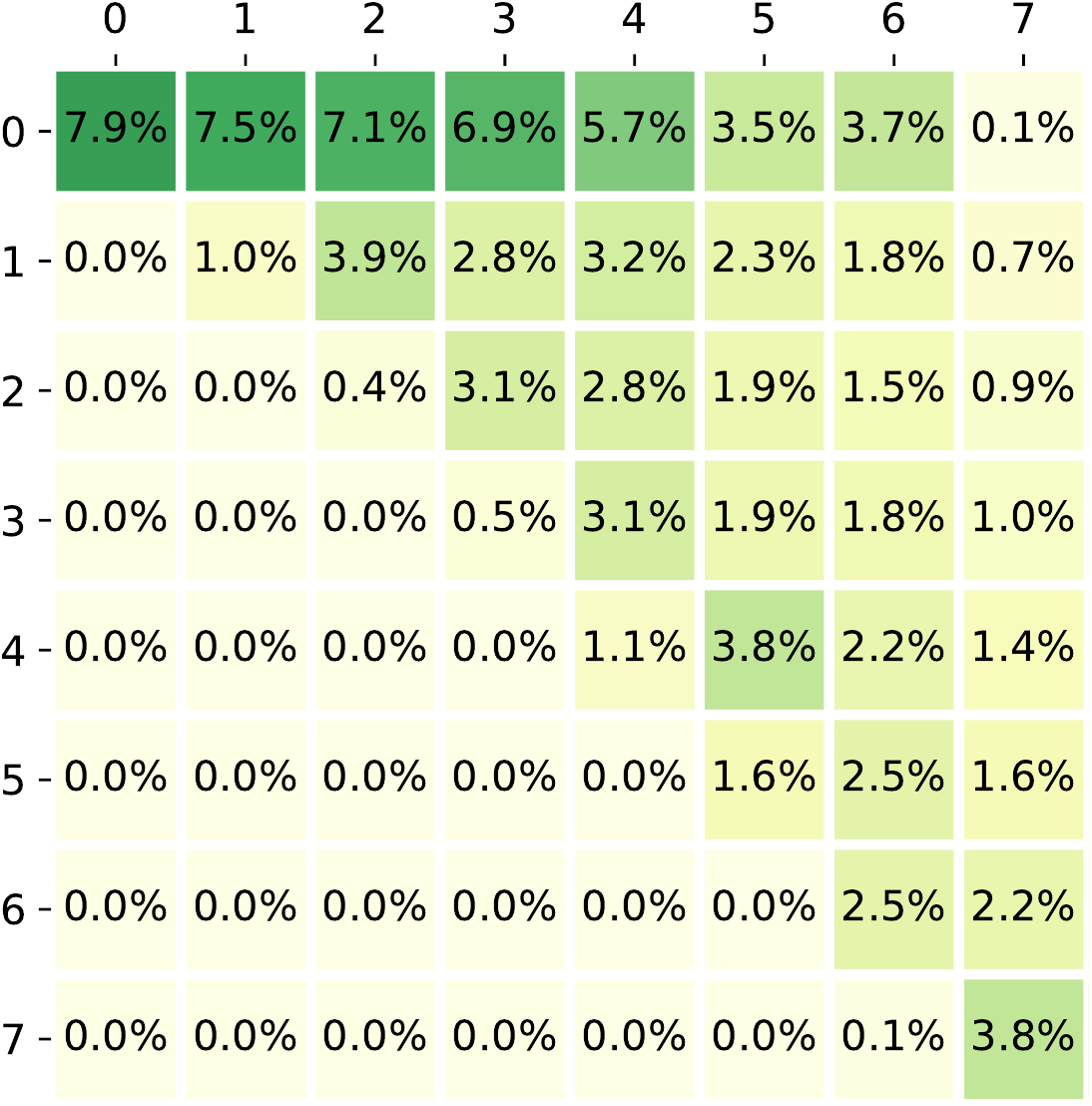}
  \label{fig:hmljnic}}
  \subfigure[UNI-LiveJournal]{\includegraphics[width=.23\linewidth]{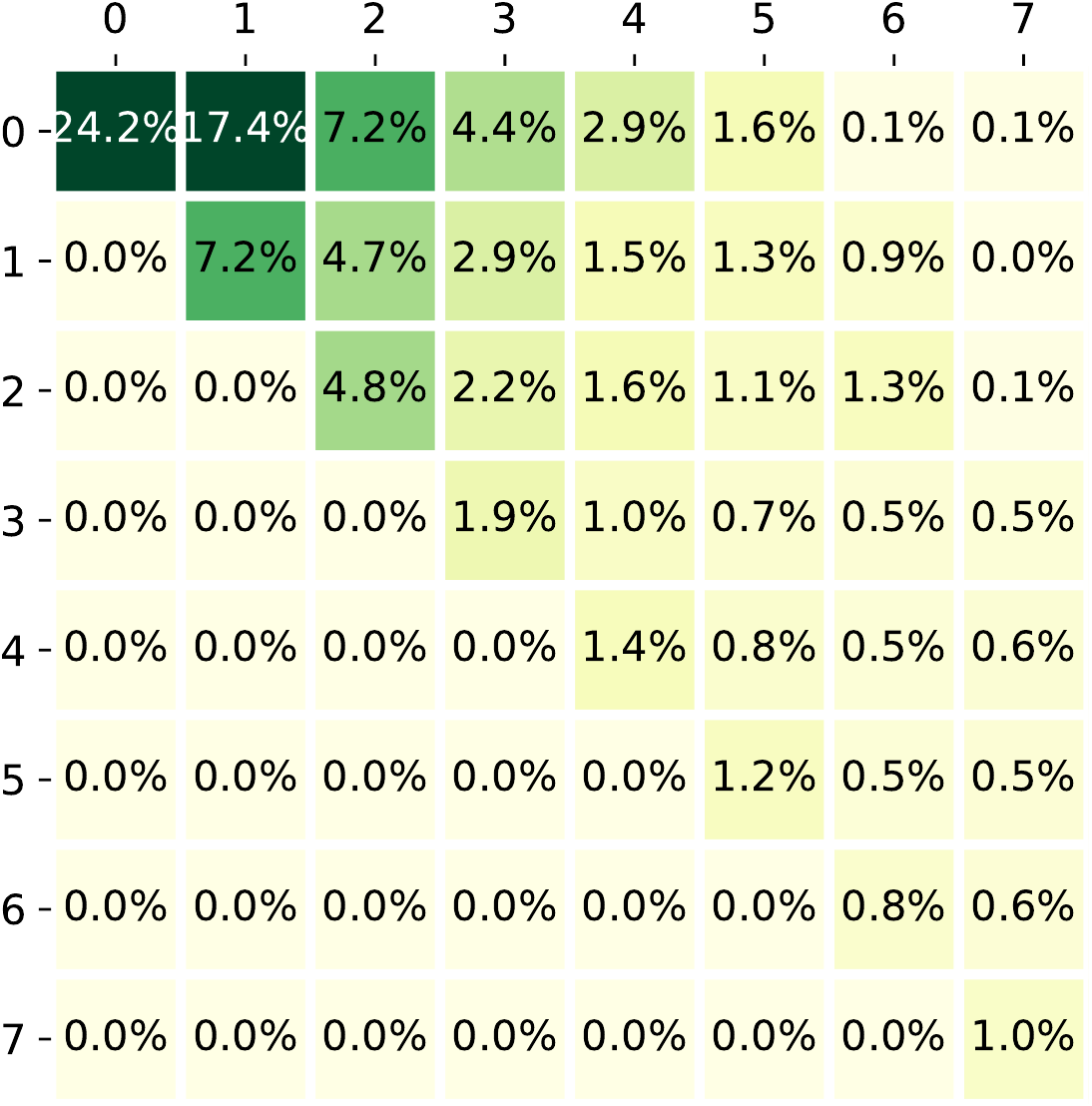}
  \label{fig:hmljuni}}
  \subfigure[PBD-LiveJournal]{\includegraphics[width=.23\linewidth]{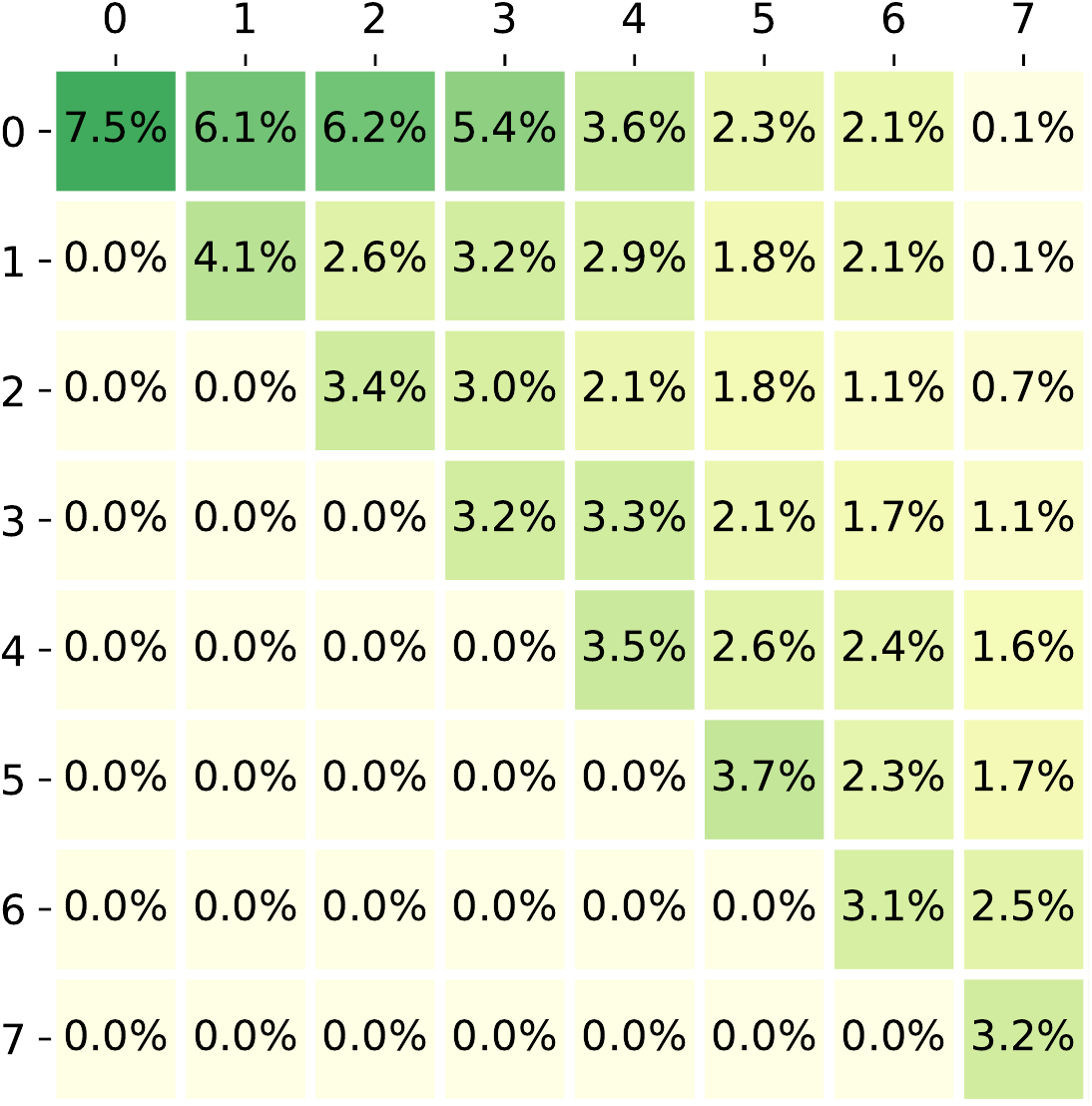}
  \label{fig:hmljpbd}}
  \subfigure[PTC-LiveJournal]{\includegraphics[width=.23\linewidth]{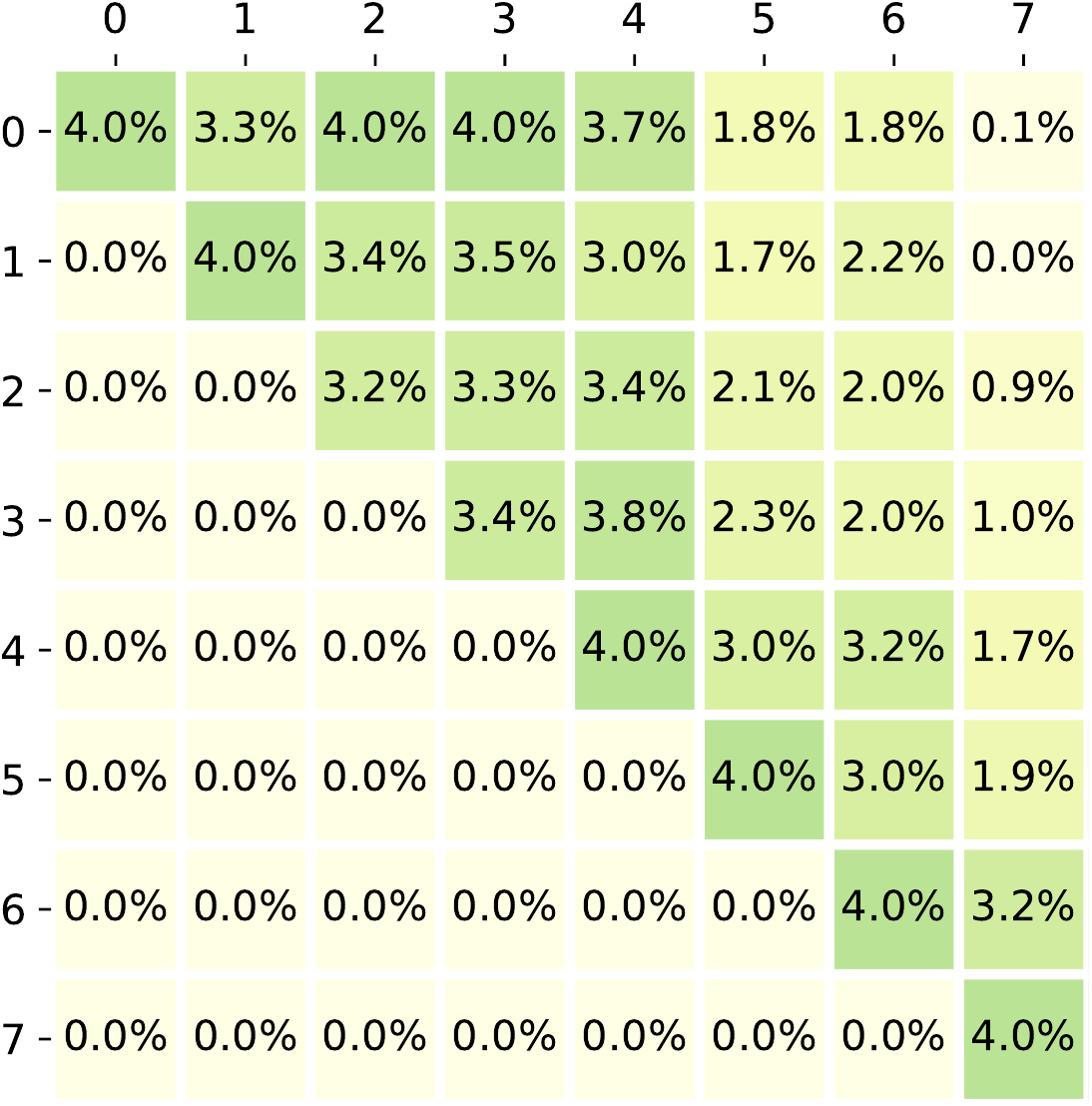}
  \label{fig:hmljptc}}

  \subfigure[NIC-Friendster]{\includegraphics[width=.23\linewidth]{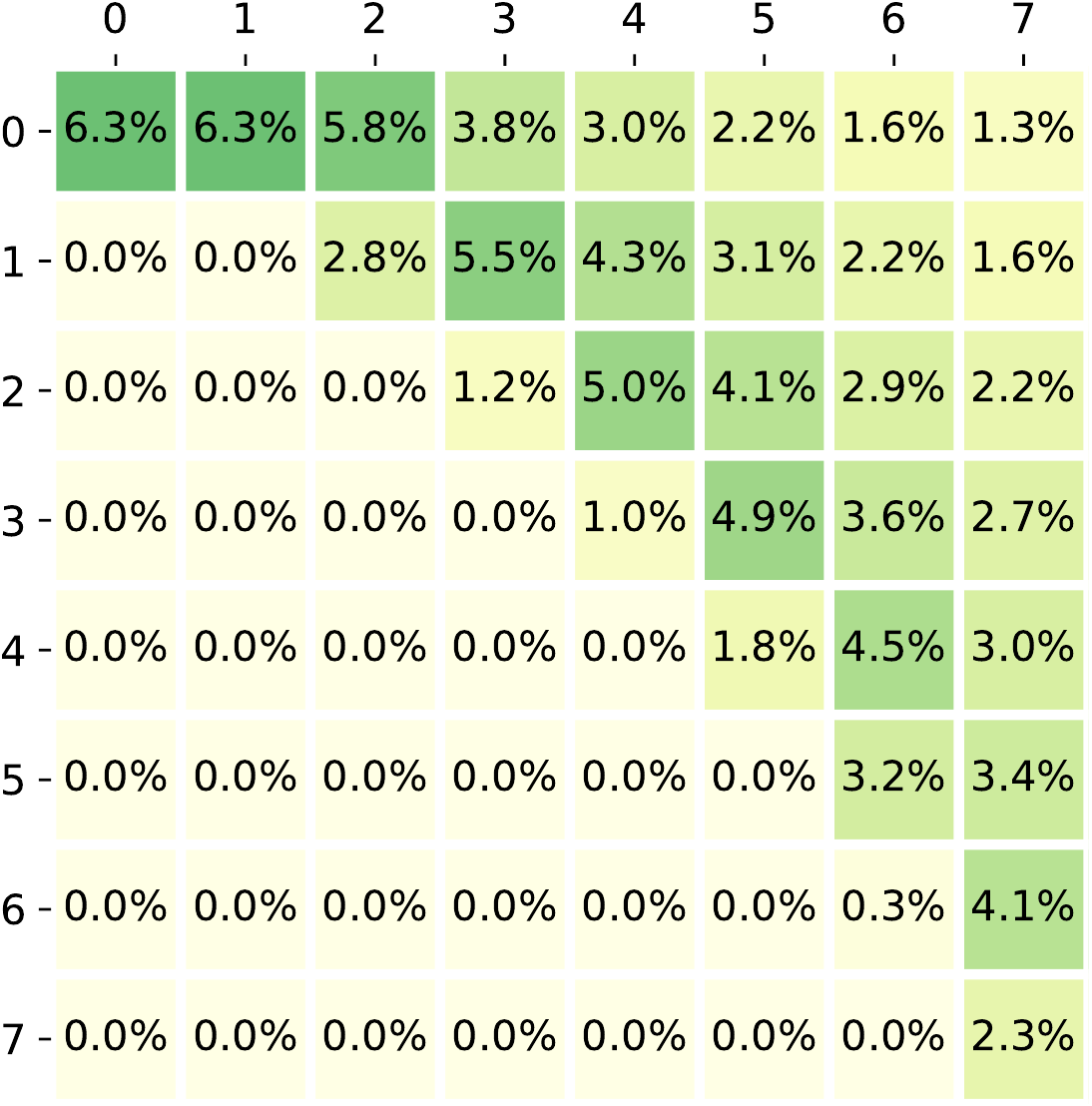}
  \label{fig:hmfrndnic}}
  \subfigure[UNI-Friendster]{\includegraphics[width=.23\linewidth]{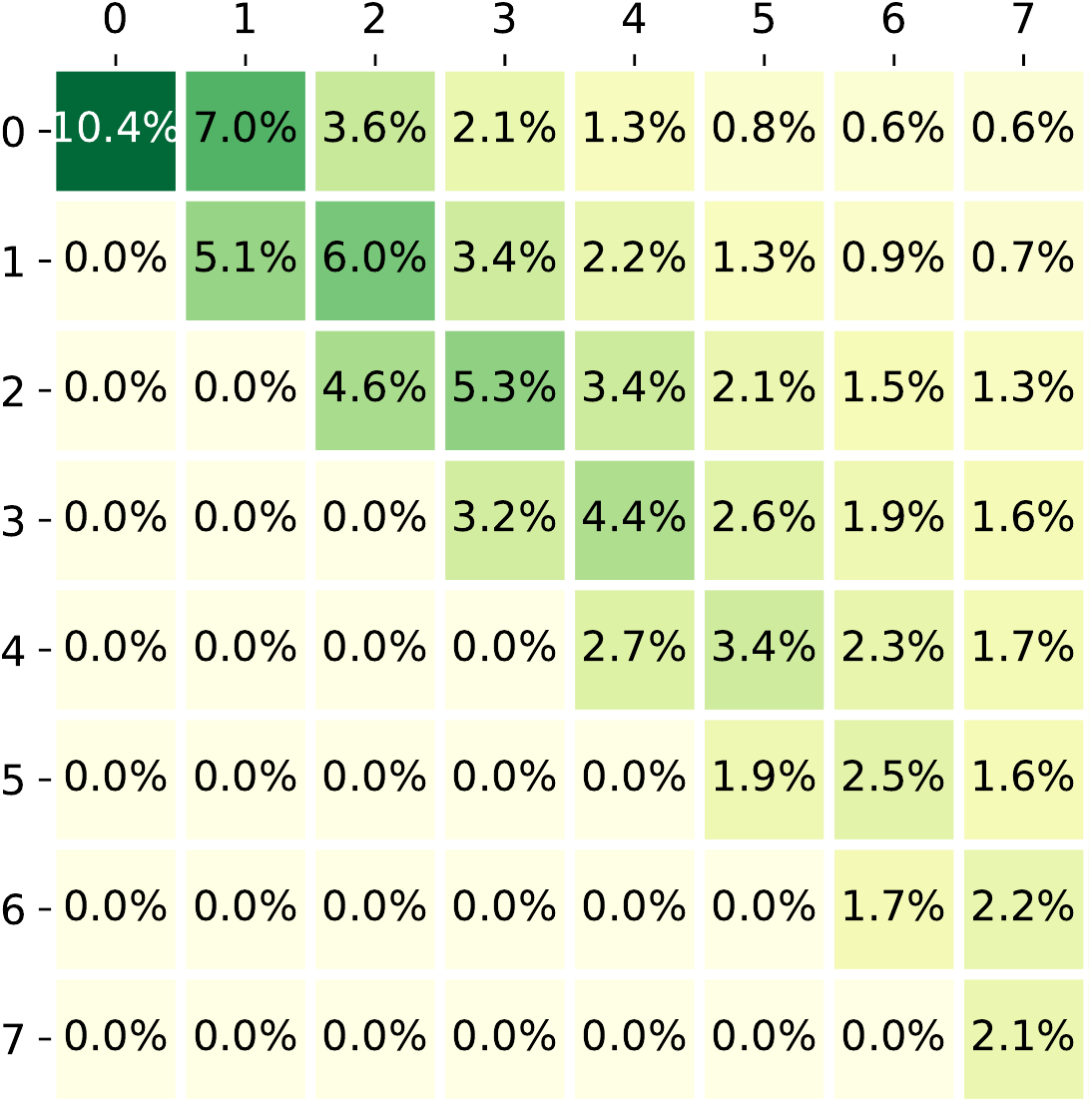}
  \label{fig:hmfrnduni}}
  \subfigure[PBD-Friendster]{\includegraphics[width=.23\linewidth]{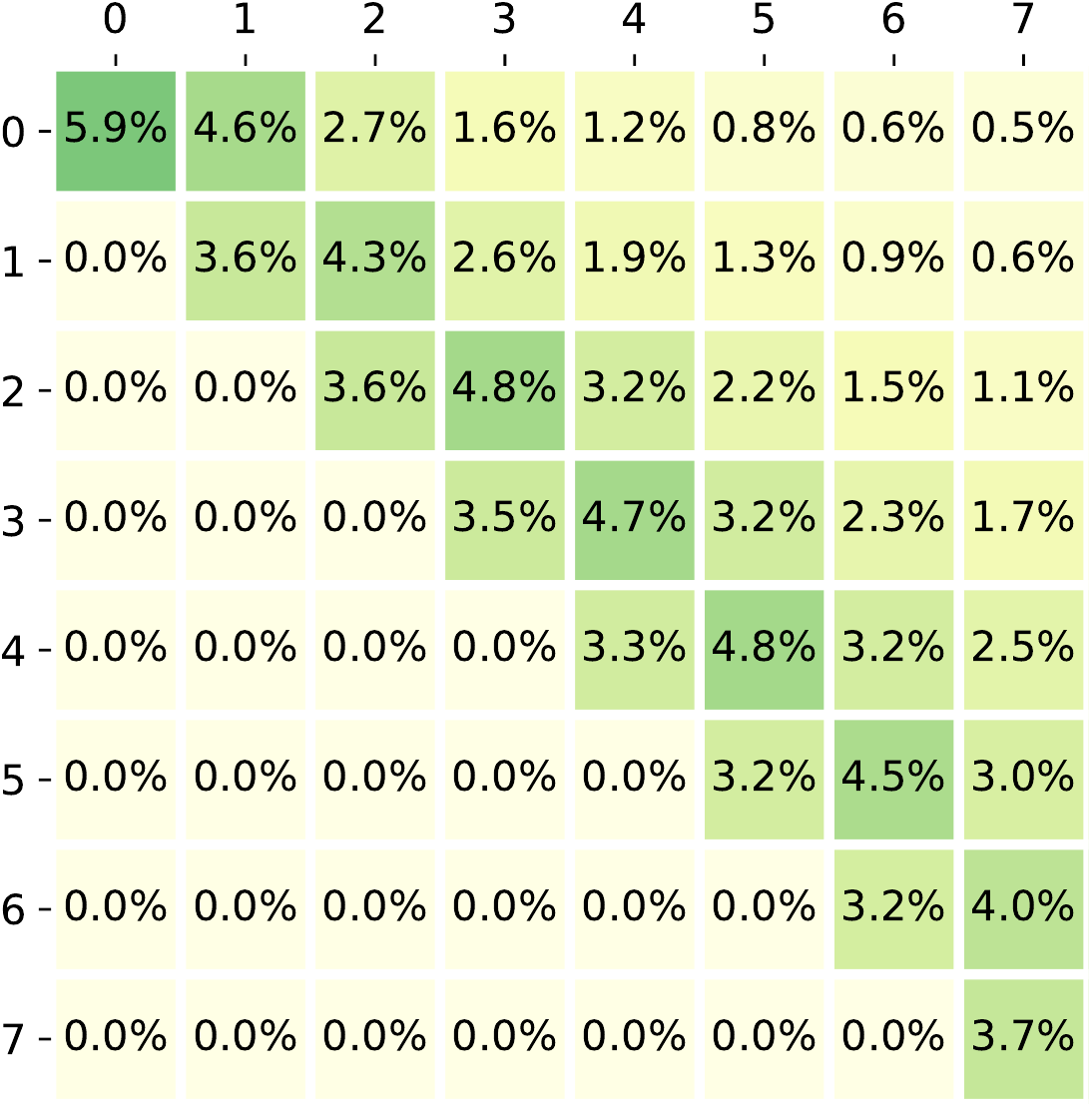}
  \label{fig:hmfrndpbd}}
  \subfigure[PTC-Friendster]{\includegraphics[width=.23\linewidth]{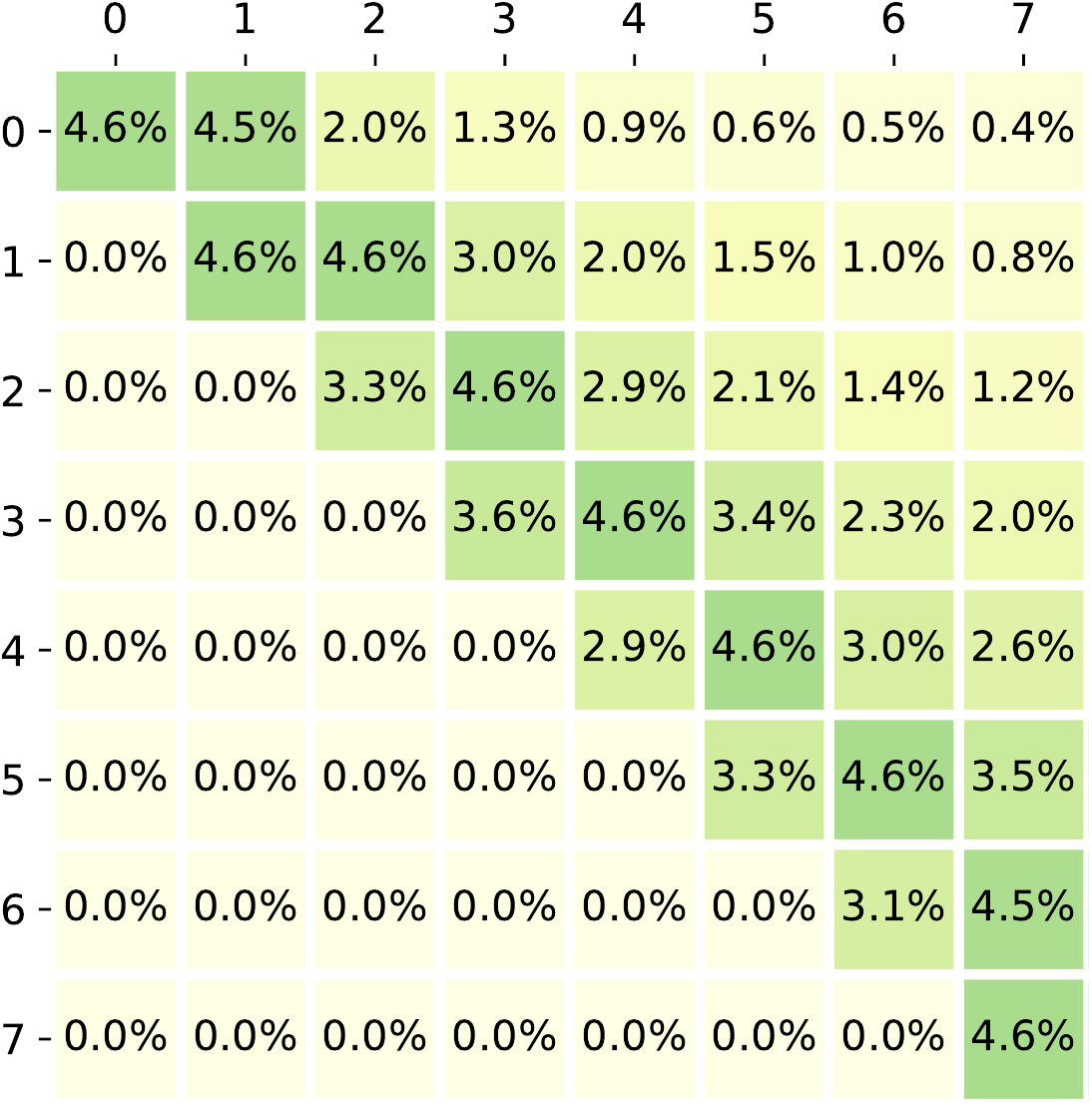}
  \label{fig:hmfrndptc}}

  \caption{Density maps of tiles in NIC, UNI, PBD and PTC based partitionings ($p=q=8$).}
  \label{fig:densitymap}
\end{figure*}

In \Cref{fig:densitymap}, we present density maps of NIC, UNI, PBD and PTC
based partitioning where $p=8$ on soc-LiveJournal1 and friendster graphs.
The same color range used in all plots and darkness of the color of a tile is
proportional to the number of nonzeros inside the tile.
In this experiment, NAT vertex ordering is used.
Percentages presented for each tile present percentage of the number of
nonzeros within a tile.
Note that in these figures tiles are mapped into a grid for a better
visualization. hence NIC algorithm outputs a rectilinear partitioning,
therefore, there can be nonzero tiles under diagonal tiles.
\Cref{fig:hmljptc} visually supports our previous
findings; PTC algorithm produces better partitioning (closely colored tiles)
and UNI gives the worst (too much variance between tile colors)
and PBD performs slightly better then NIC algorithm.

\subsection{Number of cut evaluation.}

In this experiment, we evaluate performances of BTL and PTL algorithms, proposed
for the \mNC problem.
\Cref{table:pb2} reports the number of cuts found by these algorithms
for different graphs. We choose $Z=\frac{m}{8}$ for each graph.
UNI algorithm outputs the minimum number of cuts
that can be gathered using uniform partitioning.
In \Cref{table:pb2} for each graph, the minimum number of cuts
is highlighted using a green color.
We observe that NAT vertex ordering has a lower number of cuts in general
due to the more uniform distribution of the nonzeros. On the other hand algorithms
output higher number of cuts using RCM vertex ordering,
because RCM creates denser regions while trying to make nonzeros appear near diagonal.
As expected UNI partitioning performs the worst.

\setlength{\tabcolsep}{1pt}
\begin{table}[!ht]
  \supersmallfont
    \begin{center}
      \begin{tabular}{ | l|| 
          c c c ||
          c c c ||
          c c c |
        }
        \hline

        \multirow{2}{*}{\textbf{Data Set}}  &
        \multicolumn{3}{c||}{\textbf{NAT}}  &
        \multicolumn{3}{c||}{\textbf{DEG}}  &
        \multicolumn{3}{c|}{\textbf{RCM}}  \\

        & UNI & BTL & PTL
        & UNI & BTL & PTL
        & UNI & BTL & PTL\\ \hline \hline

    cit-HepTh
    & 5 & 5 & \cellcolor{green!50} 4
    & 7 & 6 & 5
    & 6 & 6 & 5\\ \hline

    email-EuAll
    & 5 & 5 & \cellcolor{green!50} 4
    & 7 & 6 & 5
    & 6 & 6 & 5\\ \hline

    soc-Epinions1
    & 8 & 6 & \cellcolor{green!50} 4
    & 7 & 6 & 5
    & 6 & 6 & 5\\ \hline

    cit-HepPh
    & 15 & 6 & \cellcolor{green!50} 5
    & 35 & 7 & 6
    & 20 & 6 & \cellcolor{green!50} 5\\ \hline

    soc-Slashdot0811
    & 5 & 5 & \cellcolor{green!50} 4
    & 13 & 6 & 5
    & 9 & 6 & 5\\ \hline

    soc-Slashdot0902
    & 5 & 5 & \cellcolor{green!50} 4
    & 16 & 5 & 5
    & 9 & 6 & 4\\ \hline

    flickrEdges
    & 7 & 6 & \cellcolor{green!50} 5
    & 15 & \cellcolor{green!50} 5 & \cellcolor{green!50} 5
    & 8 & 6 & \cellcolor{green!50} 5\\ \hline

    amazon0312
    & \cellcolor{green!50} 5 & \cellcolor{green!50} 5 & \cellcolor{green!50} 5
    & 120 & 8 & 8
    & 25 & 8 & 7\\ \hline

    amazon0505
    & \cellcolor{green!50} 5 & \cellcolor{green!50} 5 & \cellcolor{green!50} 5
    & 71 & 6 & \cellcolor{green!50} 5
    & 48 & 6 & 6\\ \hline

    amazon0601
    & 7 & \cellcolor{green!50} 5 & \cellcolor{green!50} 5
    & 46 & \cellcolor{green!50} 5 & \cellcolor{green!50} 5
    & 21 & \cellcolor{green!50} 5 & \cellcolor{green!50} 5\\ \hline

    scale18
    & 6 & \cellcolor{green!50} 5 & \cellcolor{green!50} 5
    & 82 & 6 & 6
    & 36 & 6 & 6\\ \hline

    scale19
    & 6 & \cellcolor{green!50} 5 & \cellcolor{green!50} 5
    & 99 & 6 & \cellcolor{green!50} 5
    & 44 & 6 & 6\\ \hline

    as-Skitter
    & 6 & \cellcolor{green!50} 5 & \cellcolor{green!50} 5
    & 127 & \cellcolor{green!50} 5 & \cellcolor{green!50} 5
    & 52 & 6 & 6\\ \hline

    scale20
    & 6 &  \cellcolor{green!50} 5 &  \cellcolor{green!50} 5
    & 166 &  \cellcolor{green!50} 5 &  \cellcolor{green!50} 5
    & 57 & 6 & 6\\ \hline

    cit-Patents
    & 6 & 5 & 5
    & 86 & 5 & \cellcolor{green!50} 4
    & 34 & 6 & 5\\ \hline

    scale21
    & 14 & 6 & \cellcolor{green!50} 4
    & 63 & 5 & 5
    & 15 & 6 & 5\\ \hline

    soc-LiveJournal1
    & 7 & 6 & \cellcolor{green!50} 5
    & 50 & \cellcolor{green!50} 5 & \cellcolor{green!50} 5
    & 23 & 6 & \cellcolor{green!50} 5\\ \hline

    wb-edu
    & 7 & 6 & \cellcolor{green!50} 5
    & 50 & \cellcolor{green!50} 5 & \cellcolor{green!50} 5
    & 21 & 6 & \cellcolor{green!50} 5\\ \hline

    twitter
    & 6 & \cellcolor{green!50} 5 & \cellcolor{green!50} 5
    & 275 & 7 & 6
    & 14 & \cellcolor{green!50} 5 & \cellcolor{green!50} 5\\ \hline

    friendster
    & 10 & 8 & 8
    & 33 & 7 & \cellcolor{green!50} 5
    & 9 & 8 & 6\\ \hline

      \end{tabular}
    \caption{Number of cuts that algorithms output for the targeted load.
    Target load is $ m/8$ for each graph.}
    \label{table:pb2}

        \end{center}
\end{table}

\subsection{Execution time evaluation.}
\label{ssec:runtime}

\Cref{table:runtime} shows the execution time behavior of six algorithms; NIC, PBI, PBD, PTC, BTL,
and PTL, on the largest five graphs in our dataset listed in \Cref{table:dataset}.
We observe that, for \mNC problem both BTL and PTL algorithms' execution times are similar.
For \mLI problem, for the smaller three graphs, both PBI and PBD algorithms' execution time is
about 1.5 times slower than NIC. On largest two graphs PBD is the fastest algorithm.
Given the computational complexities of the algorithms in \Cref{tab:complexity}, as expected
PTC algorithm was the slowest of all, due to too many prefix sum lookups on the matrix.
Please note that, \Cref{tab:complexity} assumes linear lookup for a given partition vector and
require linear space. We first developed a naive implementation with such linear lookup, which
was two to three orders of magnitude slower than than other algorithms.
To decrease the lookup complexity for PTC algorithm we used a sparse prefix sum
data structure inspired from the Fenwick tree~\cite{Fenwick94-SPE}, that can query the matrix in
logarithmic time but requires $\log(n)$ times more space.
With that data structure PTC algorithm now $4$ to $43$ times slower than NIC algorithm.

\setlength{\tabcolsep}{5pt}
\begin{table}[!ht]
  \supersmallfont
    \begin{center}
      \begin{tabular}{ | l|| 
          r r r r ||
          r r |
        }
        \hline

        \multirow{2}{*}{\textbf{Data Set}}  &
        \multicolumn{4}{c||}{\textbf{mLI}}  &
        \multicolumn{2}{c|}{\textbf{mNC}}  \\

        & NIC & PBI & PBD & PTC
        & BTL & PTL \\ \hline \hline

    scale21
    & 0.9 & 1.4 & 1.6 & 17.9
    & 5.3 & 3.9 \\ \hline

    soc-LiveJournal1
    & 1.7 & 2.4 & 4.5 & 26.7
    & 245 & 269 \\ \hline

    wb-edu
    & 1.8 & 2.6 & 2.4 & 7.19
    & 14.7 & 17.1 \\ \hline

    twitter
    & 81 & 126 & 42 & 2,817
    & 190 & 186 \\ \hline

    friendster
    & 145 & 235 & 70 & 6,700
    & 245 & 269 \\ \hline

      \end{tabular}
    \caption{Execution times in seconds, of six different algorithms on the
    largest five graphs in our dataset for $8\times 8$ partitioning.}
    \label{table:runtime}

        \end{center}
\end{table}

\section{Conclusion}
\label{sec:conc}

In this work, we proposed different heuristics for symmetric
rectilinear partitioning problem and
present a thorough
experimental evaluation showing the effectiveness of the proposed
algorithms.
Even though our problem definition is more restricted, in our experiments,
we observed that our proposed algorithms give better
load-imbalance than Nicol's~\cite{Nicol94-JPDC} state-of-the-art rectilinear
partitioning algorithm in every test instances.
PTC algorithm gives the best load imbalance in the majority of the test instances
and PBD algorithm is the second-best algorithm. PBI algorithm's performance
is almost identical to NIC.

As future work, we are working on decreasing the space requirements
of our sparse prefix sum data structure and also decreasing the number of lookups that PTC
performs. In addition, we will also investigate approximation techniques, and
parallelization of the algorithm.

\section{Acknowledge}
We would like to extend our gratitude to M. Fatih Bal\.{i}n for his valuable
comments and sharing his sparse prefix sum data structure code with us.

\newpage
\bibliographystyle{siam}
\bibliography{paper}

\end{document}